\documentclass[aps, prl, reprint, nofootinbib, superscriptaddress, showpacs]{revtex4-1}
\usepackage{graphicx}
\usepackage{color}
\usepackage{mathrsfs}
\usepackage{times}
\usepackage{lineno}
\usepackage{amsmath}
\usepackage[colorlinks = true, linkcolor = blue, urlcolor  = blue, citecolor = blue, anchorcolor = blue]{hyperref}
\usepackage[textsize=footnotesize]{todonotes}

\usepackage{color}
\definecolor{oldtxtcolor}{rgb}{0.00, 0.0, 0.5}
\definecolor{newtxtcolor}{rgb}{0.00, 0.3867, 0.00}
\definecolor{newtxtcolor}{rgb}{0.00, 0.0, 1}
\definecolor{oldtxtcolor}{rgb}{1.00, 0.0, 0.00}

\def\verX{12}
\def\verO{1}
\def\verN{2}
\def\verON{12}

\ifx\verX\verO
 \newcommand { \oldtxt }[1] {{\color{oldtxtcolor}{#1}}}
 \newcommand { \newtxt }[1] {}
\fi
\ifx\verX\verN
 \newcommand { \oldtxt }[1] {}
 \newcommand { \newtxt }[1] {{\color{newtxtcolor}{#1}}}
\fi
\ifx\verX\verON
 \newcommand { \oldtxt }[1] {{\color{oldtxtcolor}{#1}}}
 \newcommand { \newtxt }[1] {{\color{newtxtcolor}{#1}}}
\fi

\usepackage{xspace}

\def\ket#1{\ensuremath{|#1\rangle}}

\def\IP#1#2{\langle#1|#2\rangle}
\def\BK#1#2#3{\langle#1|#2|#3\rangle}

\def\mol{$\mathrm{CH}_2\mathrm{OH}$\xspace}
\def\FB{\vec{F}_\mathrm{B}}
\def\Rmex{\vec{R}^{\mathrm{MECI}}_{\mathrm{O-H}}}
\def\HM{\hat{H}_0}
\def\HSO{\hat{H}_{\mathrm{SO}}}
\def\HZee{\hat{H}_{\mathrm{Z}}}
\def\Tens#1{{\mathop{#1}\limits^{\leftrightarrow}}}

\begin{document}

\title{Tracking Berry curvature effect in molecular dynamics by ultrafast magnetic x-ray scattering}
\author{Ming Zhang}
\affiliation{State Key Laboratory for Mesoscopic Physics and Collaborative Innovation Center of Quantum Matter, School of Physics, Peking University, Beijing 100871, China}
\author{Xiaoyu Mi}
\affiliation{State Key Laboratory for Mesoscopic Physics and Collaborative Innovation Center of Quantum Matter, School of Physics, Peking University, Beijing 100871, China}
\author{Linfeng Zhang}
\affiliation{State Key Laboratory for Mesoscopic Physics and Collaborative Innovation Center of Quantum Matter, School of Physics, Peking University, Beijing 100871, China}
\author{Chengyin Wu}
\affiliation{State Key Laboratory for Mesoscopic Physics and Collaborative Innovation Center of Quantum Matter, School of Physics, Peking University, Beijing 100871, China}
\affiliation{Collaborative Innovation Center of Extreme Optics, Shanxi University, Taiyuan, Shanxi 030006, China}
\affiliation{Peking University Yangtze Delta Institute of Optoelectronics, Nantong, Jiangsu 226010, China}
\author{Zheng Li}
\email{zheng.li@pku.edu.cn}
\affiliation{State Key Laboratory for Mesoscopic Physics and Collaborative Innovation Center of Quantum Matter, School of Physics, Peking University, Beijing 100871, China}
\affiliation{Collaborative Innovation Center of Extreme Optics, Shanxi University, Taiyuan, Shanxi 030006, China}
\affiliation{Peking University Yangtze Delta Institute of Optoelectronics, Nantong, Jiangsu 226010, China}

\begin{abstract}
The spin-dependent Berry force is a genuine effect of Berry curvature in molecular dynamics, which can dramatically result in spatial spin separation and change of reaction pathways.
However, the way to probe the effect of Berry force remains challenging, because the time-reversal (TR) symmetry required for opposite Berry forces conflicts with TR symmetry breaking spin alignment needed to observe the effect, and the net effect could be transient for a molecular wave packet.
We demonstrate that in molecular photodissociation, the dissociation rates can be different for molecules with opposite initial spin directions due to Berry force.
We showcase that the spatially separated spin density, which is transiently induced by Berry force as the molecular wave packet passes through conical intersection, can be reconstructed from the circular dichroism (CD) of ultrafast non-resonant magnetic x-ray scattering using free electron lasers.
\end{abstract}

\maketitle

Berry curvature~\cite{Berry392:PMP84} is one of the most fundamental properties of evolving quantum mechanical wave functions, and has far-reaching consequences in quantum systems such as topological insulators~\cite{Bernevig_book_13,Liu118:PRL17,Kane82:RMP10,Nobel89:RMP17}, quantum Hall states~\cite{Ezawa_book_08,Haldane93:PRL04,bruno93:PRL04,Xiao82:RMP10} and molecules~\cite{Mead64:RMP92,Garcia96:PRL06,Yuan362:Science18,Juanes309:Science05}.
For molecules, the Berry curvature of electronic wave function can result in the spin-dependent Berry force, which acts on the nuclei as an additional spin-dependent force apart from the gradient of potential energy surface (PES)~\cite{Subotnik151:JCP19,Culpitt155:JCP21}. 
As a consequence, molecular nuclear wave packets with opposite spin polarization experience opposite Berry forces due to the time-reversal (TR) symmetry, and the corresponding reaction pathways can be dramatically altered, which leads to spatial spin separation in a swarm of molecular dynamics trajectories~\cite{Wu12:NC21,Wu124:JPCA20}.

Though the presence of Berry force has been proposed for the molecular systems~\cite{Bian154:JCP21}, the way to observe its effect in molecular dynamics remains largely unknown.
The difficulty firstly comes from the time-reversal symmetry.
To observe the effect of Berry force, it is indispensable to align the molecular spin by an external magnetic field.
However, because the alignment magnetic field breaks time-reversal symmetry, its field strength must satisfy the condition such that the time-reversal symmetry must not be too strongly broken, otherwise the Berry forces of the spin up and down states are not nearly opposite, and as a consequence, the spin separation can be suppressed.
Besides, the spin density signal arises from spin polarized Boltzmann distribution of initial states due to Zeeman energy splitting should be suppressed, otherwise it can mix with the spin density signal originated from Berry force.
These conditions concerning TR symmetry must be resolved by a balanced magnetic field strength to enable observation of the Berry curvature effect in molecules (see Sec.~I of Supplementary Material (SM)~\cite{supplementary_info} for analysis of appropriate magnetic field strength).
Since the Berry curvature only becomes large near conical intersections, ultrafast temporal resolution may be required to capture the transient spatial spin separation induced by the Berry force within the hundreds-femtosecond characteristic time scale of molecular reaction, because the spin density signal originated from Berry force may vanish in the product states of molecules.
Moreover, because Berry force is dependent on velocities of nuclei~\cite{Subotnik151:JCP19}, the random distribution of initial velocities and molecular rotation could smear out the net effect.
Thus, in order to practically observe the net effect of Berry force, it demands proper molecular structures and reaction pathways that are distinguishable under rotational averaging, adequate magnetic field strength for spin alignment, and spatiotemporally resolved probe of the transient Berry curvature effect in the coupled spin-nuclei dynamics.

\begin{figure}
    \centering
    \includegraphics[width=8.5cm]{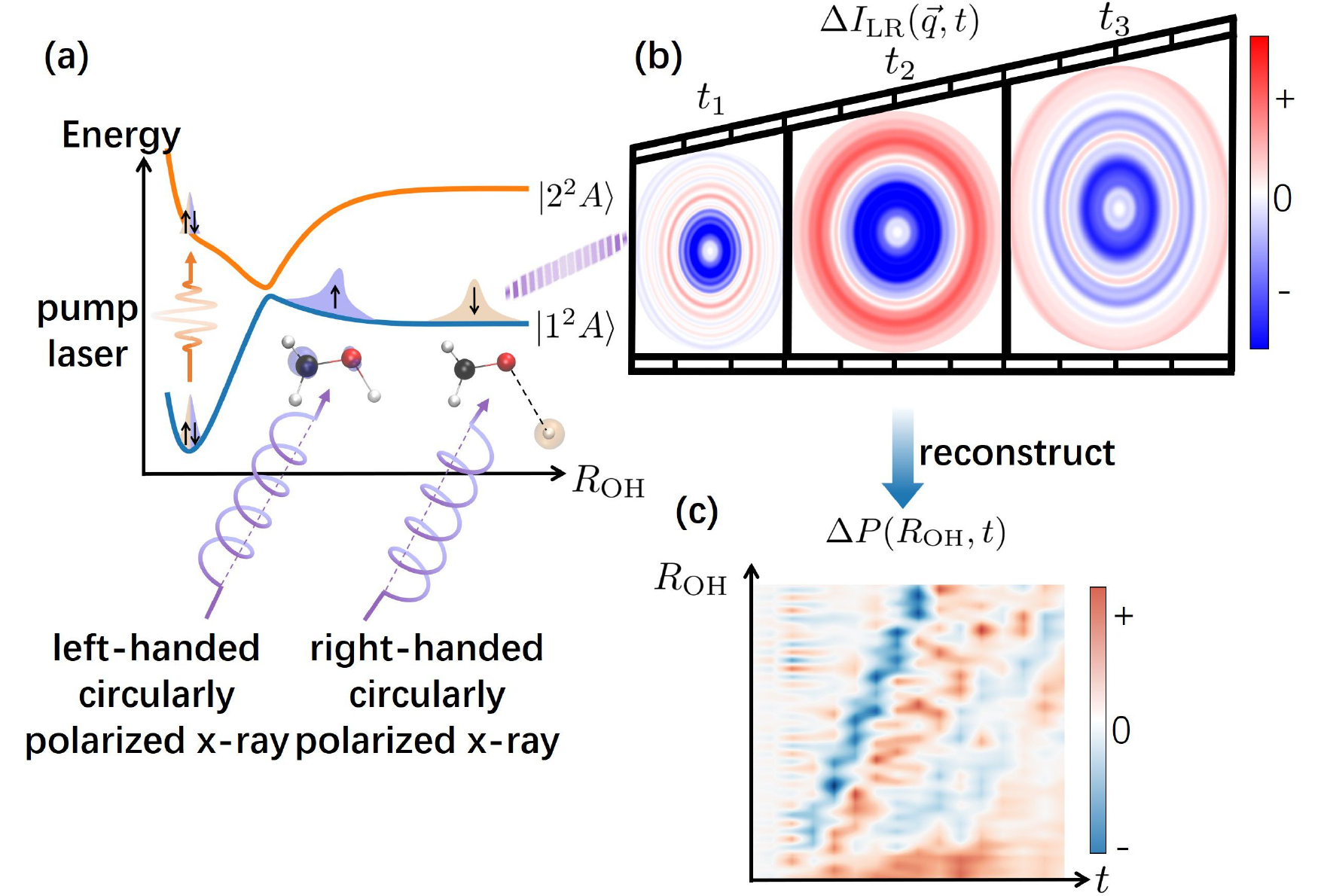}
    \caption{%
    {Schematic of Berry force induced transient spatial spin separation, which can be probed and reconstructed by the circular dichroism (CD) signal of ultrafast magnetic x-ray scattering (MXS).}
    (a) During the \mol dissociation triggered by an ultraviolet (UV) pump laser, the large Berry curvature near conical intersection leads to a spin-dependent Berry force.
    When passing through the conical intersections, the wave packet of opposite spin directions evolves to molecular geometries with different O--H distance due to opposite Berry force, namely the spin is spatially separated.
    (b) The transient spatial spin separation is detected by circular dichroism of ultrafast magnetic x-ray scattering, which is the difference of scattering intensity $\Delta I_{\mathrm{LR}}(\vec{q},t)$ between left-handed and right-handed circularly polarized x-ray.
    {(c) The Berry force induced spatial spin separation, which can be reflected by the difference of O--H distance probability density $\Delta P(R_{\mathrm{OH}},t)=P^{\uparrow}(R_{\mathrm{OH}},t)-P^{\downarrow}(R_{\mathrm{OH}},t)$ between spin up and spin down molecules, can be reconstructed from the MXS CD signal.}
    In (a), the black, red and white spheres represent C, O and H atoms in \mol, and the spin up and spin down wave packet, as well as the isosurfaces of 0.05~\AA$^{-3}$ spin density are marked by blue and orange, respectively.
    {In (b) and (c), red and blue colors represent positive and negative CD signal $\Delta I_{\mathrm{LR}}(\vec{q},t)$ and difference of O--H distance probability density $\Delta P(R_{\mathrm{OH}},t)$, respectively.}
    }
    \label{fig:schematic}
\end{figure}

In this Letter, we showcase that the Berry curvature effect of molecular electronic wave functions can be revealed by circular dichroism (CD) in ultrafast non-resonant magnetic x-ray scattering (MXS). 
As shown in Fig.~\ref{fig:schematic}, we consider the \mol photodissociation involving ground state $\ket{1^2A}$ and the first excited state $\ket{2^2A}$ (3$s$ Rydberg state)~\cite{Malbon146:JCP17,Malbon119:JPCA15}, where the Berry force acting on the molecule is conspicuously enhanced near the conical intersection.
Based on ab initio magnetic x-ray scattering calculation and molecular dynamics simulation, we demonstrate that {the spatial spin separation induced by the Berry curvature effect during \mol photodissociation, can be unambiguously probed and quantitatively reconstructed by ultrafast MXS.}
The ultrafast MXS, which is facilitated by the synchrotron radiation sources and free electron lasers (FEL)~\cite{Ohsumi1:APX16}, can resolve the instantaneous spin density via the magnetic interaction between molecules and incident x-ray photon~\cite{BookXrayScatter}.
In our scheme, hard x-ray wavelength is chosen in order to acquire the \AA{}ngstrom scale spatial resolution that is indispensable to resolve the spatial spin separation of the molecular wave packet, while it simultaneously provides higher magnetic interaction strength, which is proportional to $\frac{\hbar\omega}{mc^2}$.
We employ the circurlar dichroism of ultrafast x-ray scattering to characterize spin dynamics, because the strong charge scattering background can be eliminated by subtracting cross sections of left- and right-handed circularly polarized x-ray scattering intensities~\cite{Platzman9:PRB70,Tanaka70:PRL93,BookXrayScatter,Schuler8:npjQM23}.
As the dissociation proceeds, the spacing of diffraction fringes $\Delta q\sim\frac{1}{R_{\mathrm{OH}}}$ decreases.
Since the momentum space resolution must be finite, the circular dichroism signal practically becomes too weak to resolve, as we demonstrate numerically in Fig.~S7 in Sec.~IIIC of SM~\cite{supplementary_info}. 
In this case, 100-fs temporal resolution is required to probe the spin-nuclei coupled dissociation dynamics before the CD signal vanishes.

We begin by outlining the theory of Berry force in molecular dynamics, using the non-adiabatic photodissociation $\mathrm{CH}_2\mathrm{OH}\rightarrow\mathrm{H}+\mathrm{CH}_2\mathrm{O}$ of hydroxymethyl radical as example.
The dissociation starting from the $\ket{2^2A}$ state is triggered by an ultraviolet (UV) pump pulse of 385~nm ~\cite{Feng117:JCP02}.
The reaction proceeds in a magnetic field $\vec{B}$, which aligns the spin along the field axis. 
The Hamiltonian is
\begin{eqnarray}\label{Eq:Ham}
    \hat{H}&=&\HM+\HSO+\HZee\,,
\end{eqnarray}
where $\HM$ is the molecular Coulomb Hamiltonian including electronic and nuclear kinetic energy and their Coulomb interaction energy, $\HSO$ and $\HZee$ are the Hamiltonians for the spin-orbit coupling and Zeeman effect. 
The Berry force for the molecular dynamics (MD) simulation and the electronic wavefunction for the MXS simulation are calculated with restricted active space self-consistent field (RASSCF) method~\cite{Malmqvist90:JPC94} and cc-pVTZ basis. 
The active space consists of 5 electrons and 7 orbitals.
The Berry curvature $\Tens{\Omega}$ for the eigenstates of the electron Hamiltonian $\hat{H}$ is given by (in atomic units, au)
\begin{eqnarray}\label{Eq:BerryCurv}
{\Omega}_{I\alpha,J\beta}=-2\operatorname{Im}\IP{\nabla_{I\alpha}\psi}{\nabla_{J\beta}\psi}\,,    
\end{eqnarray}
where $I,J$ are the indices of atoms in the molecule and $\alpha,\beta$ label the molecular frame axes $x,y,z$.
From Eq.~(\ref{Eq:BerryCurv}), only the imaginary part of $\IP{\nabla_{I\alpha}\psi}{\nabla_{J\beta}\psi}$ contributes to the Berry curvature, which originates from the complex part of spin-orbit interaction Hamiltonian $\HSO$~\cite{Tao14:JPCL23}. 
Non-trivial Berry curvature imposes Berry force $\FB$ upon the nuclei apart from the gradient of the electronic PES.
Given the velocities of nuclei $\vec{v}$, the Berry force is~\cite{Subotnik151:JCP19}
\begin{eqnarray}\label{Eq:BF}
   {F}_{\mathrm{B},I\alpha}=\sum_{J\beta}{v}_{J\beta}{\Omega}_{I\alpha,J\beta}\,.
\end{eqnarray} 
$\vec{F}_{\mathrm{B}}$ is opposite for time-reversal state pair $\ket{\psi}$ and $\hat{\Theta}\ket{\psi}$ with opposite spin directions, where $\hat{\Theta}$ is the time-reversal operator.
The external magnetic field is indispensable to keep the Berry curvature effect from vanishing after averaging over initial spin orientations (see Sec.~IA of SM~\cite{supplementary_info} for symmetry analysis that shows vanishing total spin separation without initial spin alignment in a magnetic field).
We choose the magnetic field strength to be $B=0.1$~T.
In this case, the TR symmetry is not strongly broken, namely the pair of trajectories initiating from nearly degenerate time-reversal eigenstates with opposite spin directions will experience almost opposite Berry forces, and the spin polarization of initial states with $B=0.1$~T is negligibly small (In SM~\cite{supplementary_info} Sec.~IB, IC and ID, we analyze influence of magnetic field to Berry curvature effects, TR symmetry and spin polarized Boltzmann distribution).

\begin{figure}
    \centering
    \includegraphics[width=8cm]{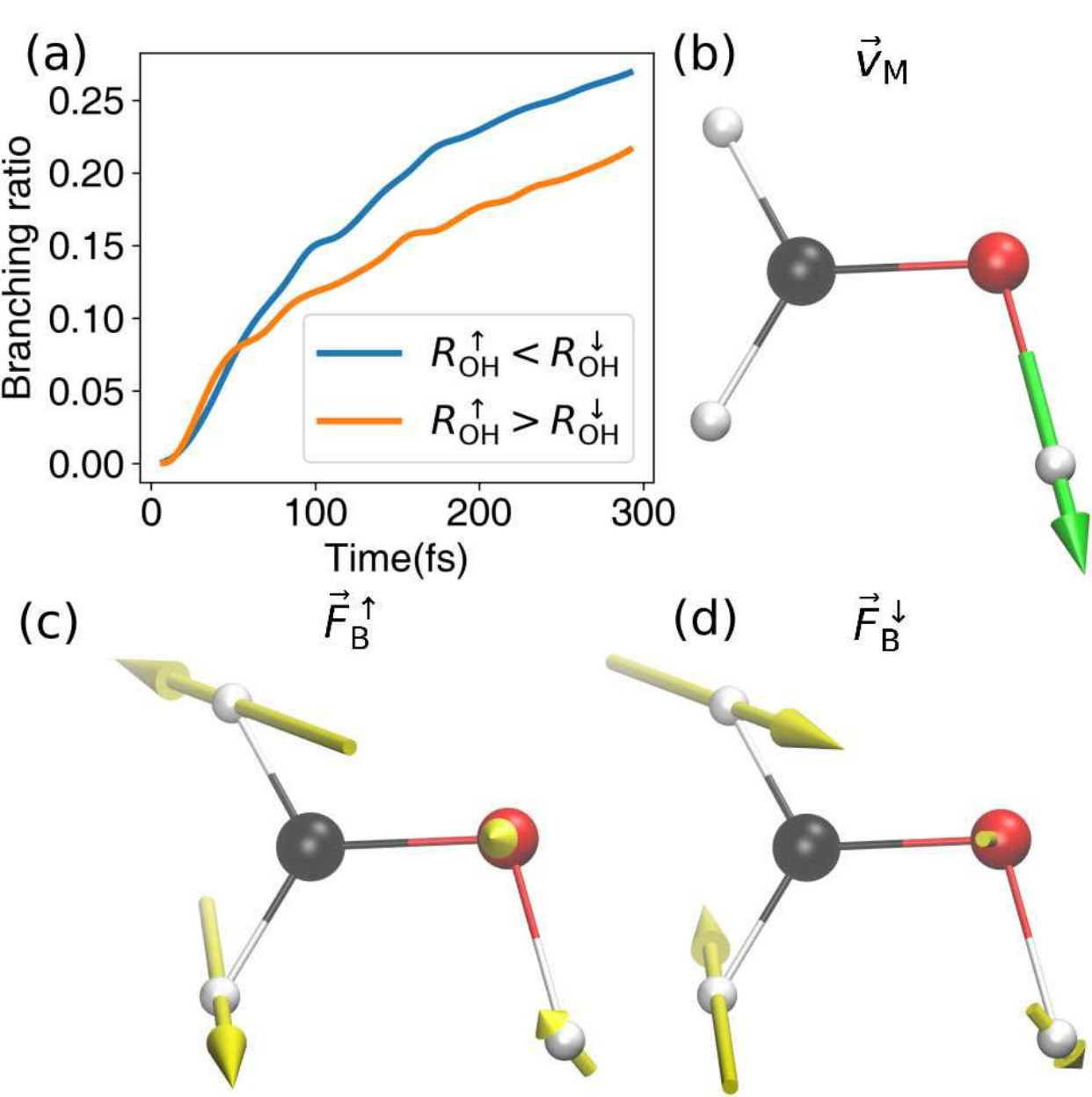}
    \caption{Molecular dynamics (MD) simulation of \mol photodissociation.
    (a) Temporal evolution for the {branching ratio} of spin separated trajectories. The dissociation rate of O--H bond of spin up trajectories is slower than spin down trajectories, which reflects the Berry force induced spatial spin separation. 
    {The branching ratios add up less than 1, because the trajectories of $|R_{\mathrm{OH}}^{\uparrow}-R_{\mathrm{OH}}^{\downarrow}|<0.1$~\AA\xspace are excluded, which represent the O--H distances difference is negligible and are not regarded as spin separated.}
    (b) The major velocity vector $\vec{v}_{\mathrm{M}}$ of MD trajectories obtained from principal component analysis (PCA) and (c) its corresponding Berry force $\FB^{\uparrow}$ for spin up state and (d) $\FB^{\downarrow}$ for spin down state at the representative conical intersection geometry $\Rmex$.
    }
    \label{fig:MD}
\end{figure}

In the photodissociation of \mol, the spin-dependent Berry forces lead to distinct MD trajectories for the molecules with opposite initial spin directions.
The MD simulation is carried out based on a multistate potential energy surfaces~\cite{Malbon146:JCP17} using modified fewest-switches surface hopping (FSSH) method~\cite{Bian18:JCTC22}, in which the Berry force is included (see Sec.~IIA of SM~\cite{supplementary_info} for modified FSSH algorithm, and Sec.~IIB for the details of molecular dynamics simulation).
The pair of MD trajectories with identical initial conditions of nuclei but opposite spin directions will experience almost opposite Berry forces, and the wave packet components with opposite spins are spatially separated, as shown in Fig.~\ref{fig:schematic}.
Fig.~\ref{fig:MD}(a) illustrates the difference of O--H bond dissociation rates for spin up and spin down trajectories, {where more trajectories in spin down state dissociates faster than their spin up counterpart.}
In order to exclusively assign the spin-dependent dissociation rate difference to Berry force, we carry out MD simulations using FSSH algorithm without adding Berry force for comparison, which shows no difference for spin up and spin down trajectories (see Fig.~S4 in Sec.~IIB of SM~\cite{supplementary_info}).

The direction of Berry force $\FB$ depends on nuclear velocities as defined in Eq.~(\ref{Eq:BF}).
However, the net effect of $\FB$ is not cancelled out due to random initial velocity directions sampled from Wigner distribution, because the MD trajectories exhibit a major velocity vector $\vec{v}_{\mathrm{M}}$ along the O--H dissociation coordinate as shown in Fig.~\ref{fig:MD}(b).
$\vec{v}_{\mathrm{M}}$ is obtained from principal component analysis (PCA) of MD trajectories~\cite{Sittel141:JCP14}.
The corresponding Berry force $\FB=\vec{v}_{\mathrm{M}}\cdot\Tens{\Omega}$ is shown in Fig.~\ref{fig:MD}(c)(d), where the Berry curvature is calculated using the representative minimal energy conical intersection geometry $\Rmex$~\cite{Malbon146:JCP17} between $\ket{1^2A}$ and $\ket{2^2A}$ states (see Fig.~S1 and Fig.~S3 in Sec.~I of SM~\cite{supplementary_info} for details of $\Rmex$ geometry and Berry curvature calculation). 
As shown in Fig.~\ref{fig:MD}(c)(d), the direction of Berry force for spin up state $\FB^{\uparrow}$ and spin down state $\FB^{\downarrow}$ hinders and enhances O--H dissociation, respectively.

Berry force leads to the spatial spin separation of nuclear wave packet, thus the spin density for spin up trajectories differ from that of spin down trajectories.
The spatial spin separation is reflected by the circular dichroism signal of ultrafast non-resonant magnetic x-ray scattering.
Define the Fourier transformed spin density operator $\hat{\vec{s}}(\vec{q})=\sum_{jj'}e^{-i\vec{q}\cdot{(\vec{r}_j-\vec{r}_{j'})}}\hat{\vec{\sigma}}_{j'}$, where $\vec{r}_j$ and $\hat{\vec{\sigma}}_j$ are position and Pauli matrices vectors of the $j$-th electron, $\vec{q}$ is momentum transfer of scattered x-ray photon, and $\BK{\psi}{\hat{\vec{s}}(\vec{q})}{\psi}$ is spin density of state $\ket{\psi}$ in the Fourier space.
The circular dichroism differential cross section of MXS signal $\Delta\frac{d\sigma}{d\Omega}$ is calculated by averaging over the swarm of MD trajectories (see Sec.~IIIA of SM~\cite{supplementary_info} for detailed formulae of ultrafast MXS CD signal)
\begin{eqnarray}\label{Eq:MXSCD}
 \langle\Delta\frac{d\sigma}{d\Omega}(\vec{q},t)\rangle=\frac{\alpha^6}{N_{\mathrm{tr}}}\sum_i\operatorname{Re}\left[\vec{s}^{\,\uparrow\downarrow}_{i}(\vec{q},t)\cdot\vec{D}(\vec{q})\right]\,,   
\end{eqnarray}
where $\alpha$ is the fine-structure constant, $\vec{D}(\vec{q})$ is the polarization factor, $N_{\mathrm{tr}}$ is the number of MD trajectories.
$\vec{s}^{\,\uparrow\downarrow}_{i}(\vec{q},t)=p^{\uparrow}\BK{\psi^{\uparrow}(\vec{R}^{\uparrow}_i,t)}{\hat{\vec{s}}(\vec{q})}{\psi^{\uparrow}(\vec{R}^{\uparrow}_i,t)}+p^{\downarrow}\BK{\psi^{\downarrow}(\vec{R}^{\downarrow}_i,t)}{\hat{\vec{s}}(\vec{q})}{\psi^{\downarrow}(\vec{R}^{\downarrow}_i,t)}$ is the Fourier space spin density contributed from the $i$-th spin up and spin down MD trajectories with same initial geometry and velocity, where $p^{\uparrow}$ and $p^{\downarrow}$ are Boltzmann distributions of initial spin states.
The spin density signal by magnetic x-ray scattering unambiguously reveals the Berry curvature effect in \mol dissociation dynamics, because nonzero spin density signal $\vec{s}^{\,\uparrow\downarrow}_i\neq 0$ implies spatial spin separation, namely, $\vec{R}^{\uparrow}_i(t)\neq\vec{R}^{\downarrow}_i(t)$ from the same initial conditions.
The spin density contributed purely by the magnetic field $B=0.1$~T aside from Berry force is negligible (see Sec.~IIIC of SM~\cite{supplementary_info} for estimation of net spin density caused by initial Boltzmann distribution of spin states).
\begin{figure*}
    \centering
    \includegraphics[width=15cm]{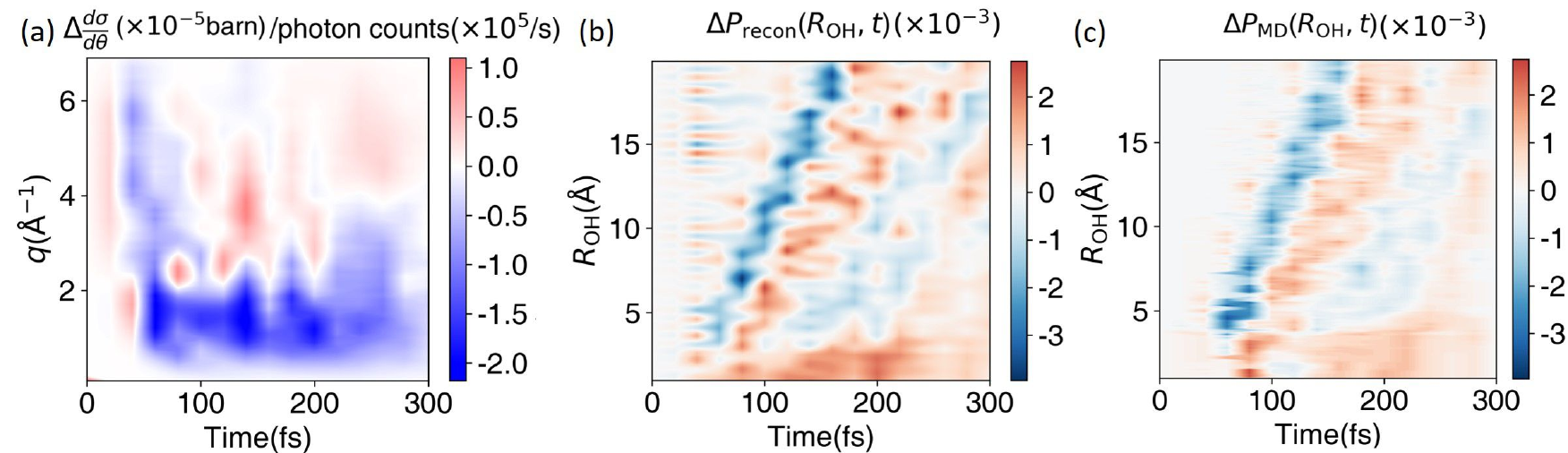}
    \caption{%
    Simulated ultrafast magnetic x-ray scattering circular dichroism signal and {the difference of spin dependent probability density of O--H distance} for the swarm of molecular dynamics trajectories.
    (a) Temporal evolution of the MD averaged MXS CD cross section $\langle\Delta\frac{d\sigma}{d\theta}(q,t)\rangle$ and estimated scattered photon numbers per second with our chosen free electron laser (FEL) parameters.
    The $10^{-5}$ barn cross section corresponds to $10^{5}$ scattered photons per second.
    {(b) From the ultrafast MXS CD signal, we can reconstruct the temporal evolution for the difference of O--H distance probability density between spin up and spin down trajectories $\Delta P_{\mathrm{recon}}(R_{\mathrm{OH}},t)$, which reflects the Berry force induced spatial spin separation.}
    (c) The difference of spin dependent probability density $\Delta P_{\mathrm{MD}}(R_{\mathrm{OH}},t)$ calculated from MD trajectories, which is consistent with (b).}
    \label{fig:MXS_MD}
\end{figure*}

Based on the MD trajectories, we simulate the ultrafast non-resonant MXS CD signal, and the result is shown in Fig.~\ref{fig:MXS_MD}(a) (see Sec.~IIIB of SM~\cite{supplementary_info} for details of ab initio MXS simulation).
We assume isotropic angular distribution for molecular rotational wavepacket.
The wave vector of incident x-ray and the magnetic field is along $Z$ axis in the lab frame, and the wavelength of x-ray is 0.5~\AA.
As shown in Fig.~\ref{fig:MXS_MD}(a), the intensity of MXS CD signal increases within 100~fs, as the Berry force induces spatial spin separation.
{The intensity MXS CD signal descends with O--H dissociation, making MXS CD effectively a transient phenomena.}
The descent of MXS CD signal intensity can be interpreted as the spacing of diffraction fringes $\Delta q\sim\frac{1}{R_{\mathrm{OH}}}$ decreases with the dissociation of \mol, which goes beyond the resolution along with O--H dissociation and leads to the descent of MXS CD signal 
(see Fig.~S7 in Sec.~IIIC of SM~\cite{supplementary_info} for the change of MXS CD signal intensity with O--H distance $R_{\mathrm{OH}}$).
The scattered x-ray photon counts of MXS CD signal, as shown in Fig.~\ref{fig:MXS_MD}(a), are calculated from cross section under feasible experimental conditions of free electron laser (FEL)~\cite{Emma4:NP10,OShea292:Science01}, by assuming photon number per pulse $N_{\gamma}=10^{12}$, molecular gas density $\rho=10^{16}~\mathrm{cm}^{-3}$, the diameter of molecular gas $L=1~\mathrm{cm}$ and the repetition rate $\nu=1~\mathrm{MHz}$ (see Sec.~IIIE of SM~\cite{supplementary_info} for the estimate of scattered photon number using realistic FEL parameters).
Practically, the left- and right-handed circularly polarized scattering signals should be normalized before subtraction, in order to cancel the intensity fluctuations of FEL x-ray pulses in either the self-amplified spontaneous emission (SASE)~\cite{Emma4:NP10} or the seeded regime~\cite{Giannessi108:PRL12}.

In order to quantitatively retrieve the Berry force induced spatial spin density evolution in the ultrafast photodissociation dynamics of $\mathrm{CH}_2\mathrm{OH}$ we established an analytical model for the MXS CD signal, which is analogous to the independent atom model (IAM) in the theory of molecular diffraction ~\cite{Yangjie18:Sci64}.
The model assumes that the spin density is localized, which becomes rigorous in the dissociation limit (see sec.~IIID of SM~\cite{supplementary_info} for the model of MXS CD signal and Fig.~S8 for validity of the model).
Define the spin-dependent probability density of O--H distance $P^\uparrow(R_{\mathrm{OH}},t)$ and $P^\downarrow(R_{\mathrm{OH}},t)$ for spin up and down MD trajectories, respectively, and their difference $\Delta P(R_{\mathrm{OH}},t)=P^\uparrow(R_{\mathrm{OH}},t)-P^\downarrow(R_{\mathrm{OH}},t)$ indicates the spatial spin separation caused by Berry force.
The spin separated probability density can be reconstructed from MXS CD signal by (see Sec.~IIID of SM~\cite{supplementary_info} for the detailed derivation of Eq.~\ref{Eq:PrRt})
\begin{eqnarray}\label{Eq:PrRt}
    \Delta P_{\mathrm{recon}}(R_{\mathrm{OH}},t)=R_{\mathrm{OH}}\int dq  \frac{\sin(qR_{\mathrm{OH}})}{A(q)} \langle\Delta \frac{d\sigma}{d\theta}(q,t)\rangle\,,
\end{eqnarray}
where $A(q)=\frac{N-1}{Nq}\pi^2\alpha^6 D_Z(q)  f_{\mathrm{H}}(q)f_{X}(q)$, $N$ is the number of total electrons in \mol, $D_Z(q)$ is the $Z$-component of polarization factor, $f_{\mathrm{H}}(q)$ is the form factor of H atom, and $f_{X}(q)$ represents the form factor of CH$_2$O moiety, which is regarded as a pseudoatom and the angular dependence of $f_{X}(q)$ on the direction of momentum transfer vector $\vec{q}$ is averaged.
The spin separated probability density reconstructed from MXS CD signal $\Delta P_{\mathrm{recon}}(R_{\mathrm{OH}},t)$ and the result of MD trajectories $\Delta P_{\mathrm{MD}}(R_{\mathrm{OH}},t)$ are shown in Fig.~\ref{fig:MXS_MD}(b) and (c), which exhibit consistent behaviors {that for more trajectories, the O--H bond dissociation rates of spin down states are faster than the spin up counterpart}.

To summarize, we studied the spin-dependent photodissociation of \mol, which serves as a typical molecular dynamics system exhibiting non-trivial Berry curvature effect.
As a genuine effect of the Berry curvature, the Berry force near conical intersections separates nuclear wave packet in states with opposite spin directions. 
Besides, we demonstrate that spatial spin separation induced by Berry force can be reconstructed from the ultrafast non-resonant circular dichroism signal of magnetic x-ray scattering circular, {which serves as the imaging of spin dynamics in molecule.}
Our work paves the way to the direct observation and quantitative measurement of Berry curvature effect in the molecular systems, and has far-reaching implications to the study of {ultrafast geometric effects in molecules}.

\begin{acknowledgements}
We thank Zunqi Li for helpful discussions.
This work has been supported by National Natural Science Foundation of China (Nos.~12234002, 12174009, 92250303), and Beijing Natural Science Foundation (No.~Z220008).
\end{acknowledgements}


%

\end{document}




\title{Supplementary Material for \\
Tracking Berry curvature effect in molecular dynamics by ultrafast magnetic x-ray scattering}
\author{Ming Zhang}
\affiliation{State Key Laboratory for Mesoscopic Physics and Collaborative Innovation Center of Quantum Matter, School of Physics, Peking University, Beijing 100871, China}
\author{Xiaoyu Mi}
\affiliation{State Key Laboratory for Mesoscopic Physics and Collaborative Innovation Center of Quantum Matter, School of Physics, Peking University, Beijing 100871, China}
\author{Linfeng Zhang}
\affiliation{State Key Laboratory for Mesoscopic Physics and Collaborative Innovation Center of Quantum Matter, School of Physics, Peking University, Beijing 100871, China}
\author{Chengyin Wu}
\affiliation{State Key Laboratory for Mesoscopic Physics and Collaborative Innovation Center of Quantum Matter, School of Physics, Peking University, Beijing 100871, China}
\affiliation{Collaborative Innovation Center of Extreme Optics, Shanxi University, Taiyuan, Shanxi 030006, China}
\affiliation{Peking University Yangtze Delta Institute of Optoelectronics, Nantong, Jiangsu 226010, China}
\author{Zheng Li}
\email{zheng.li@pku.edu.cn}
\affiliation{State Key Laboratory for Mesoscopic Physics and Collaborative Innovation Center of Quantum Matter, School of Physics, Peking University, Beijing 100871, China}
\affiliation{Collaborative Innovation Center of Extreme Optics, Shanxi University, Taiyuan, Shanxi 030006, China}
\affiliation{Peking University Yangtze Delta Institute of Optoelectronics, Nantong, Jiangsu 226010, China}

\maketitle

\section{The effect of external magnetic field}
In the main text, we emphasize that the external magnetic field is indispensable to the observation of the net Berry curvature effect, which is non-vanishing only when the initial molecular spin directions are aligned.
Moreover, the magnitude of external magnetic field $B$ is required to satisfy:

1. $B$ should not be too strong, such that the spin polarization caused by Boltzmann distribution of initial states due to Zeeman energy splitting, which can also contribute to the spin density signal, is small enough compared with the signal originated from Berry force.

2. $B$ should not be too strong that the time-reversal symmetry is not severely broken, otherwise the Berry forces for the spin up and down states are not nearly opposite, which can reduce the spin separation effect.

In this section we analyze the effect of magnetic field of $B=0.1$~T in our scheme.

\subsection{Necessity of initial spin alignment in magnetic field}
In the absence of external magnetic field, the total spin separation effect vanishes, because after averaging for all spin polarization directions, the Fourier transformed total spin density is
\begin{eqnarray}\label{eq:zero_Sq}
    \vec{s}_{\mathrm{total}}(\vec{q})&=&\int_{\mathrm{SO(3)}} d \textbf{R} \BK{\hat{U}(\textbf{R})\psi}{\hat{\vec{s}}(\vec{q})}{\hat{U}(\textbf{R})\psi}\\\nonumber
    &=&\int_{\mathrm{SO(3)}} d \textbf{R} \BK{\psi}{\hat{U}^\dagger(\textbf{R})\hat{\vec{s}}(\vec{q})\hat{U}(\textbf{R})}{\psi}= \int_{\mathrm{SO(3)}} d \textbf{R} \BK{\psi}{\textbf{R}\hat{\vec{s}}(\vec{q})}{\psi}=0\,,
\end{eqnarray}
where $\hat{U}(\textbf{R}(\theta,\vec{n}))=e^{-i\theta\vec{n}\cdot\hat{\vec{\sigma}}/2}$ is the representation of 3D rotation operator $\textbf{R}(\theta,\vec{n})$, $\vec{\sigma}$ is the vector of Pauli matrices, $\hat{\vec{s}}(\vec{q})=\sum_{jj'}e^{-i\vec{q}\cdot{(\vec{r}_j-\vec{r}_{j'})}}\hat{\vec{\sigma}}_{j'}$ is the Fourier transformed spin density operator, $\textbf{R}$ ranges over all possible rotation matrices, and we used the relation of Pauli matrices $\hat{U}^\dagger(\textbf{R})\hat{\vec{\sigma}}\hat{U}(\textbf{R})=\textbf{R}\hat{\vec{\sigma}}$.
Thus, Eq.~\ref{eq:zero_Sq} implies vanishing magnetic x-ray scattering signal in the absence of magnetic field.

\subsection{Estimate of energy scales of relevant interactions}\label{sec:ene_level}
\begin{figure}
    \centering
    \includegraphics[width=10cm]{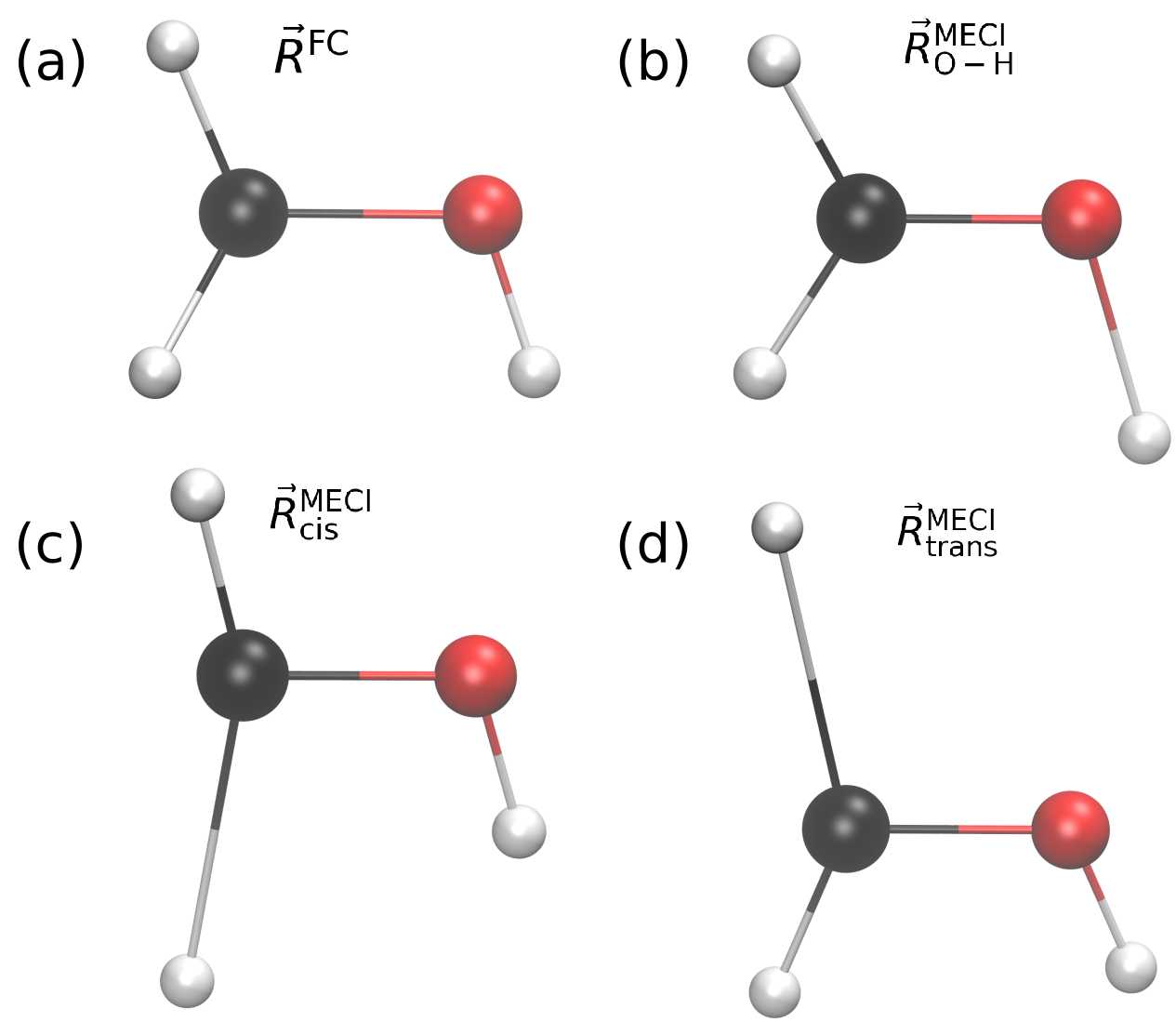}
    \caption{Representative geometries in \mol dissociation dynamics.
    The coordinates for these geometries are provided in Ref.~\cite{Malbon146:JCP17}.
    (a) Franck-Condon (FC) geometry $\RFC$, namely the equilibrium geometry in the ground state.
    (b) $\Rmex$, (c) $\Rmexcis$ and (d) $\Rmextrans$ are local minimal energy conical intersections (MECI) between $\ket{1^2A}$ and $\ket{2^2A}$ states, which can lead to H$+$CH$_2$O, H$+$cis-HCOH and H$+$trans-HCOH dissociation channels, respectively.
    Black, red and white spheres represent C, O and H atoms, respectively.
    }
    \label{fig:Geom_CI}
\end{figure}

We investigate the Berry curvature and Berry force for the representative geometries in \mol dissociation~\cite{Malbon146:JCP17}, including the Franck-Condon (FC) geometry $\RFC$, which is the equilibrium geometry in the ground state, and several local minimal energy conical intersections (MECI) between $\ket{1^2A}$ and $\ket{2^2A}$ states.
The corresponding molecular structures are shown in Fig.~\ref{fig:Geom_CI}. 
Different MECIs in Fig.~\ref{fig:Geom_CI}(b)-(d) correspond to different \mol dissociation channels, and
$\mathrm{CH}_2\mathrm{OH}\rightarrow\mathrm{H}+\mathrm{CH}_2\mathrm{O}$
is the dominant photodissociation channel from initial state $\ket{2^2A}$ ~\cite{Malbon146:JCP17,Malbon119:JPCA15}.

\begin{figure}
    \centering
    \includegraphics[width=15cm]{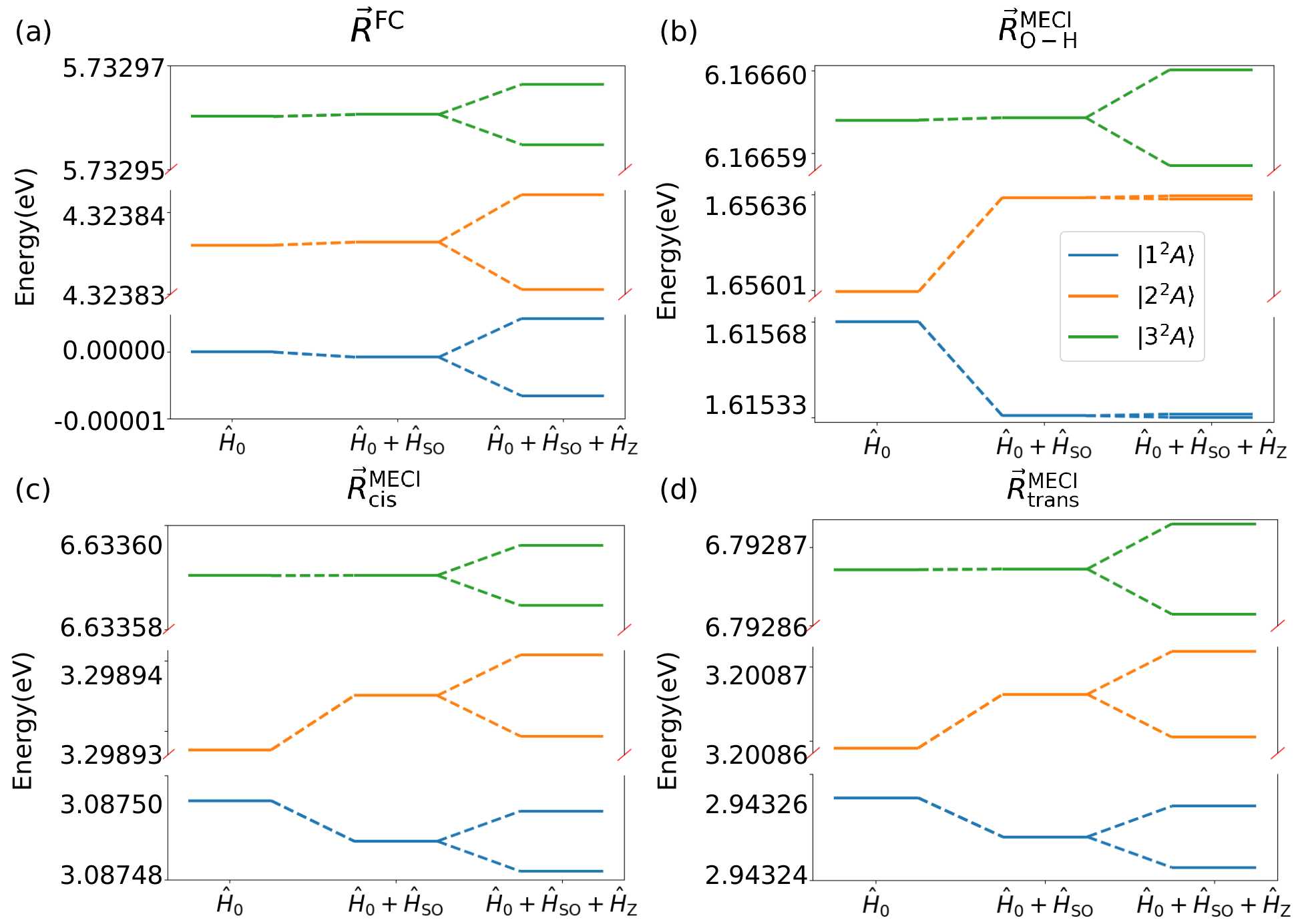}
    \caption{Energy levels for (a) FC geometry and (b)-(d) representative minimal energy conical intersections (MECI) geometries.
    Energy eigenvalues for $\HM$, $\HM+\HSO$ and total Hamiltonian $\HM+\HSO+\HZee$ defined in Eq.~\ref{Eq:Ham} are calculated for each geometry.
    The Zeeman energy splitting $\Delta E_Z$ is in the order of $10^{-5}$~eV for the magnetic field $B=0.1$~T.
    }
    \label{fig:Ene_diagram}
\end{figure}
The total Hamiltonian, as given in Eq.~1 of the main text, is
\begin{eqnarray}\label{Eq:Ham}
    \hat{H}&=&\HM+\HSO+\HZee\,.
\end{eqnarray}
$\HSO$ is Breit-Pauli form spin-orbit coupling Hamiltonian~\cite{Tao14:JPCL23,Abegg30:MP75}
\begin{eqnarray}
    \HSO=-\frac{\alpha^2}{2}\sum_{i,I}\frac{Z_I}{|\vec{r}_{iI}|^3} (\vec{r}_{iI}\times\vec{p_i})\cdot \vec{s}_i\,,
\end{eqnarray}
where $\alpha=1/137$ is the fine-structure constant, $Z_I$ is the atomic number of $I$-th nucleus, $\vec{p}_i$ and $\vec{s}_i$ are the momentum and spin for the $i$-th electron, $\vec{r}_{iI}$ is the position vector from electron $i$ to nucleus $I$.
The Zeeman Hamiltonian under magnetic field $B$ along $Z$ axis in lab frame is
\begin{eqnarray}
    \HZee&=&\frac{\alpha}{2}B(\hat{L}_Z+2\hat{S}_Z)\,,
\end{eqnarray}
where $\hat{L}_Z$ and $\hat{S}_Z$ are the projections of orbital and spin angular momentum operator along $Z$ axis, respectively.
Fig.~\ref{fig:Ene_diagram} shows the energy levels of $\HM$, $\HM+\HSO$ and total Hamiltonian $\HM+\HSO+\HZee$, respectively.
The adiabatic state energies, the matrix elements of spin-orbit coupling and orbital angular momentum $\hat{L}_Z$ are calculated on the RASSCF(5,7)/cc-pvtz level using Molpro package~\cite{MOLPRO2019}.

As shown in Fig.~\ref{fig:Ene_diagram}, both $\HM$ and $\HSO$ are time-reversal symmetric and the energy levels remain Kramers degenerate.
The eigenstates of $\HM$ are the direct product of adiabatic spatial state $\ket{\phi_k}$ and spin state $\ket{s}$ $(s=\uparrow,\downarrow)$.
Due to time-reversal symmetry we have
\begin{eqnarray}
    \BK{\phi_k}{\hat{L}}{\phi_k}&=&(\BK{\phi_k}{\hat{\Theta})(\hat{\Theta}^\dagger\hat{L}\hat{\Theta})(\hat{\Theta}^\dagger}{\phi_k})=-\BK{\phi_k}{\hat{L}}{\phi_k}\,,
\end{eqnarray}
thus $\BK{\phi_k}{\hat{L}}{\phi_k}=0$, where $\hat{\Theta}$ is the time-reversal operator.
This also explains why $\HSO$ does not induced additional splitting for spin up and spin down states.
Because the current generated from orbital motion for each spatial state $\ket{\phi_k}$ is zero, which do not give an effective magnetic field acting on electronic spin.
The Zeeman energy splitting $\Delta E_Z$ is in the order of $10^{-5}$~eV , which is much smaller compared to the energy differences between different electronic states.

\subsection{The change of Berry curvature with magnetic field}
In this section, we demonstrate that the magnetic field has negligible effect to modify the intrinsic Berry curvature of electronic wavefunction of molecules.
We show that the time-reversal symmetry is not severely broken in the magnetic field of 0.1~T, so the nearly degenerate time-reversal eigenstates with opposite spin directions in the magnetic field will experience almost opposite Berry forces.
\begin{figure}
    \centering
    \includegraphics[width=15cm]{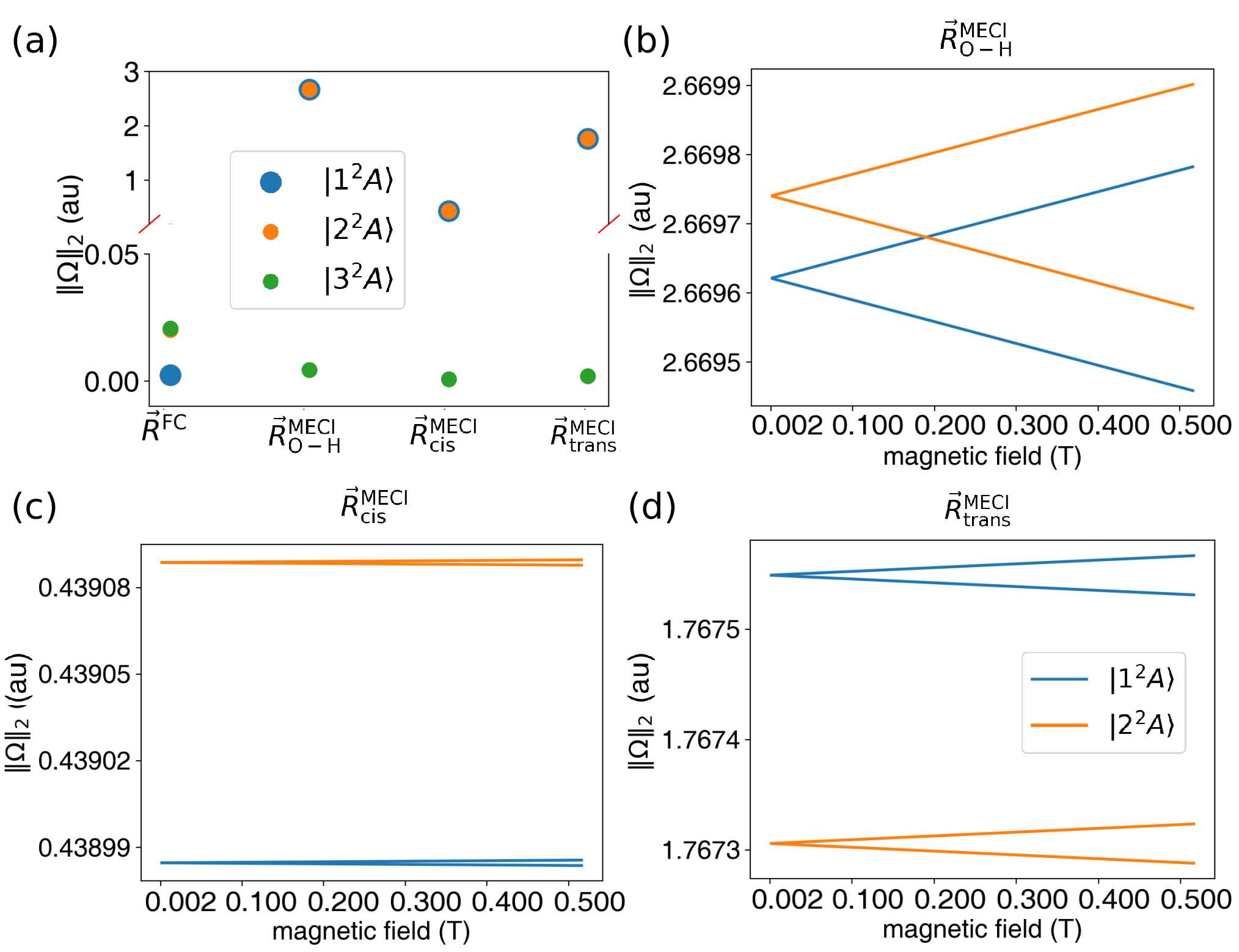}
    \caption{%
    Berry curvature for FC geometry and the representative MECIs.
    (a) The 2-norm of Berry curvature $\Vert\Omega\Vert_2$ for $\RFC$ and three representative MECIs between $\ket{1^2A}$ and $\ket{2^2A}$ states in the magnetic field of 0.1~T.
    For all the three MECIs, the $\Vert\Omega\Vert_2$ value of $\ket{1^2A}$ and $\ket{2^2A}$ states are nearly identical. For $\RFC$ and of $\ket{3^2A}$ state of the MECIs, $\Vert\Omega\Vert_2<0.005$~a.u .
    (b)-(d) The change of $\Vert\Omega\Vert_2$ with magnetic field intensity $B$ for $\Rmex$, $\Rmexcis$ and $\Rmextrans$, respectively.
    When $B$ approaches zero, the Berry curvatures are nearly opposite for the pair of almost time-reversal degenerate states, so their norms are almost identical.
    When $B=0.1$~T, $\Vert\Omega\Vert_2$ splits. However, Berry curvatures for the pair of nearly time-reversal degenerate states, which are split by Zeeman energy, are also almost opposite within $0.01\%$ relative difference.
    The minimal magnetic field for the calculation is $B=0.002$~T.
    }
    \label{fig:BerryCurv}
\end{figure}

The Berry curvature $\Tens{\Omega}$ for state $\ket{\psi}$ is defined in Eq.~2 of the main text
\begin{eqnarray}\label{Eq:BerryCurv}
{\Omega}_{I\alpha,J\beta}=-2\operatorname{Im}\IP{\nabla_{I\alpha}\psi}{\nabla_{J\beta}\psi}\,,
\end{eqnarray}
where $I,J$ are the indexes of atom in the molecule, $\alpha,\beta$ label the molecular frame directions $x,y,z$.
The Berry curvature of state $\ket{\psi}$ is opposite for its time-reversal state $\ket{\tilde{\psi}}=\hat{\Theta}\ket{\psi}$, where $\hat{\Theta}$ is the time-reversal operator, because
\begin{eqnarray}
\Tens{\Omega}(\ket{\tilde{\psi}})=-2\operatorname{Im}\IP{\nabla_{I\alpha}\hat{\Theta}\psi}{\nabla_{J\beta}\hat{\Theta}\psi}=-2\operatorname{Im}\IP{\nabla_{I\alpha}\psi}{\nabla_{J\beta}\psi}^*=-\Tens{\Omega}(\ket{\psi})\,,
\end{eqnarray}
due to the antiunitary property of $\hat{\Theta}$~\cite{Wu124:JPCA20}.

We use finite differences method to numerically calculate the Berry curvature in the vicinity of FC and representative MECI geometries~\cite{Culpitt155:JCP21}
\begin{eqnarray}
    {\Omega}_{I\alpha,J\beta}(\ket{\psi(\vec{R})})&=&-2\operatorname{Im}\IP{\frac{\psi(\vec{R}+\vec{\delta}_{I\alpha})-\psi(\vec{R}-\vec{\delta}_{I\alpha})}{2\delta_{I\alpha}}}{\frac{\psi(\vec{R}+\vec{\delta}_{J\beta})-\psi(\vec{R}-\vec{\delta}_{J\beta})}{2\delta_{J\beta}}}\\\nonumber
    &=&-\frac{1}{2\delta_{I\alpha}\delta_{J\beta}} \operatorname{Im}
    (S_{\vec{\delta}_{I\alpha},\vec{\delta}_{J\beta}}-
    S_{-\vec{\delta}_{I\alpha},\vec{\delta}_{J\beta}}-
    S_{\vec{\delta}_{I\alpha},-\vec{\delta}_{J\beta}}+
    S_{-\vec{\delta}_{I\alpha},-\vec{\delta}_{J\beta}})\,,
\end{eqnarray}
where $\vec{\delta}_{I\alpha}$ is a small displacement for the $I$-th atom and $\alpha$-th component of nuclear geometry vector $\vec{R}$, and $S_{\pm\vec{\delta}_{I\alpha},\pm\vec{\delta}_{J\beta}}=\IP{\psi(\vec{R}\pm\vec{\delta}_{I\alpha})}{\psi(\vec{R}\pm\vec{\delta}_{J\beta})}$ is the overlap of electronic wavefunction at different nuclear geometries.
We calculate $S_{\pm\vec{\delta}_{I\alpha},\pm\vec{\delta}_{J\beta}}$ using Molcas package~\cite{MOLCAS} on the same calculation level RASSCF(5,7)/cc-pvtz as in Sec.~\ref{sec:ene_level}.

The Berry curvature effects can be significantly enhanced due to non-adiabatic coupling~\cite{Wu12:NC21}.
The numerical results for the 2-norm of Berry curvature $\Vert\Omega\Vert_2$ for $\RFC$ and representative MECIs $\Rmex$, $\Rmexcis$ and $\Rmextrans$ are shown in Fig.~\ref{fig:BerryCurv}.
For all the three MECIs between $\ket{1^2A}$ and $\ket{2^2A}$ states, the $\Vert\Omega\Vert_2$ value for the non-adiabatically coupled $\ket{1^2A}$ and $\ket{2^2A}$ states are much larger compared to that of $\ket{3^2A}$ state and $\RFC$ geometry.
In the existence of weak magnetic field, time-reversal degenerate states are split by Zeeman energy.
However, as shown in Fig.~\ref{fig:BerryCurv}, Berry curvature for the pair of states $\ket{\psi}$ and $\ket{\tilde{\psi}}$ are almost opposite.
The deviation caused by magnetic field can be estimated by the relative difference
\begin{eqnarray}
\frac{\Vert\Omega(\ket{\tilde{\psi}})\Vert_2-\Vert\Omega(\ket{{\psi}})\Vert_2}{\Vert\Omega(\ket{{\psi}})\Vert_2}
\end{eqnarray}
which is less than $0.01\%$  when $B=0.1$~T for all three MECIs shown in Fig.~\ref{fig:BerryCurv}.

\subsection{Spin polarized Boltzmann distribution}
In this section, we estimate the spin polarized Boltzmann distribution resulted from Zeeman energy splitting and spin alignment dynamics in the magnetic field.
As defined in the main text, $p^{\uparrow}$ and $p^{\downarrow}$ are Boltzmann distributions of spin up and spin down initial states.
In the absence of magnetic field $p^{\uparrow}=p^{\downarrow}$.
For $B=0.1$~T and the temperature $T=300$~K, we have the Zeeman energy splitting $\Delta E_Z=\alpha B=0.1/1720\times 1/137~\mathrm{au}=4.25\times 10^{-7}~\mathrm{au}$, and $k_{\mathrm{B}}T=1.38\times 10^{-23}~\mathrm{J}/\mathrm{K}\times 300~\mathrm{K}=9.51 \times 10^{-4}~\mathrm{au}$, where $k_{\mathrm{B}}$ is the Boltzmann constant, so
\begin{eqnarray}\label{Eq:Boltz}
    |p^{\uparrow}-p^{\downarrow}|&=&\frac{1-e^{-\Delta E_Z/k_{\mathrm{B}}T}}{1+e^{-\Delta E_Z/k_{\mathrm{B}}T}}\approx\frac{\Delta E_Z}{2k_{\mathrm{B}}T}=\frac{4.25\times 10^{-7}~\mathrm{au}}{2\times9.51 \times 10^{-4}~\mathrm{au}}\approx 2\times10^{-4}\,.
\end{eqnarray}
Thus, the effect of spin polarized Boltzmann distribution to magnetic x-ray scattering signal is much smaller than the signal corresponding to Berry curvature effects (see Sec.~\ref{sec:MXS_noise} for the estimate of magnetic x-ray scattering signal intensity caused by spin polarized Boltzmann distribution).

\section{Molecular dynamics of \mol photodissociation}
We performed \mol photodissociation molecular dynamics (MD) simulation using modified fewest-switches surface hopping (FSSH) algorithm to incorporate the Berry force effect~\cite{Bian18:JCTC22}.
In this section, we present the details of modified FSSH algorithm and MD simulation.

\subsection{Modified FSSH algorithm including Berry force}
The electronic state is expanded in the direct product state between adiabatic spatial state $\ket{\phi_k(\vec{R}(t))}$ at nuclear geometry $\vec{R}(t)$ and spin state $\ket{s}$ $(s=\uparrow,\downarrow)$ as
\begin{eqnarray}
    \ket{\psi(t)}=\sum_{k,s} c_{k,s}(t)\ket{\phi_k(\vec{R}(t)),s}
\end{eqnarray}
where $c_{k,s}(t)$ is the coefficient.
The electronic state is propagated by Schr\"odinger equation as in standard FSSH algorithm
\begin{eqnarray}
    \dot{c}_{k,s}&=&-i\sum_{j,s'}\BK{\phi_k,s}{\hat{H}_{\mathrm{el}}}{\phi_j,s'}c_{j,s'}-\sum_j{\vec{v}}\cdot\BK{\phi_k}{\nabla}{\phi_j}c_{j,s}\,,
\end{eqnarray}
where $\hat{H}_{\mathrm{el}}$ is the electronic Hamiltonian including adiabatic potential, $\HSO$ and $\HZee$ in Eq.~\ref{Eq:Ham}, ${\vec{v}}$ is the nuclear velocity and $\BK{\phi_k}{\nabla}{\phi_j}$ is the derivative coupling operator.
Equivalently, the electronic state can be propagated in the adiabatic state representation, where the matrix element of $\hat{H}_{\mathrm{el}}$ is diagonalized.

The nuclear degrees of freedom is classically propagated along one ``active" adiabatic state~\cite{Culpitt155:JCP21} denoted as $\ket{\psi_\lambda}$ by
\begin{eqnarray}
	\dot{\vec{p}}=-\nabla E_\lambda+\FB+\vec{F}_{\mathrm{L}}\,,
\end{eqnarray}
where $\vec{p}$ is momentum of nuclei, $E_\lambda$ and $\FB$ are energy and Berry force of  $\ket{\psi_\lambda}$ state, respectively.
$\vec{F}_{\mathrm{L}}$ is the Lorentz force and $\vec{F}_{\mathrm{L},I}=Z_{I}\vec{v}_I\times\vec{B}$ for the $I$-th nucleus where $Z_{I}$ is the atomic number.
For each MD time step, the trajectory on active surface $\ket{\psi_\lambda}$ randomly hops to another surface $\ket{\psi_{\lambda'}}$ with transition probability~\cite{Bian18:JCTC22}
\begin{eqnarray}
	g_{\lambda\rightarrow\lambda'}=\mathrm{max}[2\operatorname{Re}\left(\vec{v}\cdot\BK{\psi_\lambda}{\nabla}{\psi_{\lambda'}}\frac{\rho_{\lambda'\lambda}}{\rho_{\lambda\lambda}}\right)\Delta t,0]\,,
\end{eqnarray}
where $\rho_{\lambda'\lambda}$ is the electronic state density matrix element and $\Delta t$ is the time step in MD.

\subsection{Molecular dynamics details}
For the molecular dynamics simulations, the analytical potential energy surface (PES) established in Ref.~\cite{Malbon146:JCP17} is used, which provides the adiabatic potential energy and derivative coupling for each geometry.
The Lorentz force arising from the external magnetic field $B=0.1$~T is much smaller compared to Berry force~\cite{Tao14:JPCL23},
\begin{eqnarray}
    \frac{|\vec{F}_{\mathrm{L}}|}{|\vec{F}_{\mathrm{B}}|}=\frac{|q\alpha \vec{v}\times\vec{B}|}{|\vec{v}\cdot\Tens{\Omega}|}\approx\frac{q\alpha B}{\Vert\Omega\Vert_2}\approx\frac{1/137\times 0.1/1720~\mathrm{au}}{1~\mathrm{au}}\approx 4\times10^{-7}\,,
\end{eqnarray}
and also much smaller than the gradient of potential energy surfaces (PES)
\begin{eqnarray}
    \frac{|\vec{F}_{\mathrm{L}}|}{|\nabla V|}\approx\frac{q\alpha v B}{|\nabla V|}\approx\frac{1/137\times 10^{-3}\times0.1/1720~\mathrm{au}}{0.1~\mathrm{au}}\approx 4\times10^{-9}\,,
\end{eqnarray}
so $\vec{F}_{\mathrm{L}}$ can be ignored in MD.

\begin{figure}
    \centering
    \includegraphics[width=7cm]{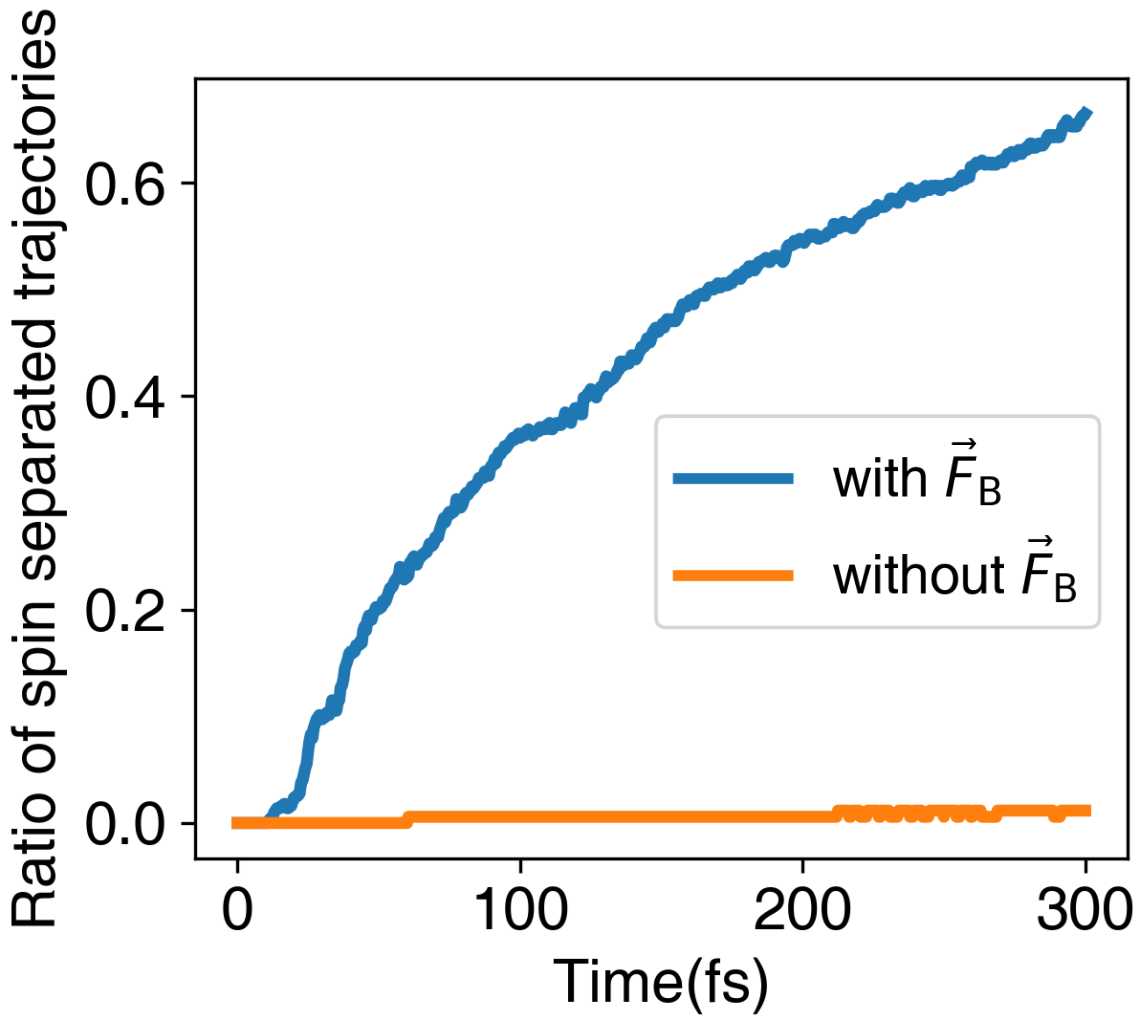}
    \caption{%
    The branching ratio of spin separated molecular dynamics (MD) trajectories.
    The blue and orange line corresponds to MD simulations with and without Berry force, respectively.
    Each pair of trajectories $\vec{R}_i^{\uparrow}(t)$ and $\vec{R}_i^{\downarrow}(t)$ with identical phase space initial conditions are classified as the spin separated configuration if $\Vert\vec{R}_i^{\uparrow}(t)-\vec{R}_i^{\downarrow}(t)\Vert>0.1~${\AA}.
    Only $<2$~\% trajectories are spin separated in the absence of Berry force, which can be ignored compared to 66.8~\% spin separation with Berry force.
    }
    \label{fig:NoBerry}
\end{figure}

We calculated $\sim 700$ pairs of spin up and spin down MD trajectories.
The initial conditions of nuclear geometries and velocities are sampled from Wigner distribution.
In order to guarantee that the spin-dependent MD results exclusively originate from Berry curvature effect, all the MD parameters, including the phase space initial conditions and the random number seed for surface hopping simulation of each pair of spin up and spin down trajectories, are identical.
For comparison, we also performed MD simulation using standard FSSH algorithm without adding $\FB$, which shows almost no differences for spin up and spin down trajectories as shown in Fig.~\ref{fig:NoBerry}.
The tiny spin separation in Fig.~\ref{fig:NoBerry} is the consequence of Zeeman energy splitting, which lifts the degeneracy of time-reversal eigenstates with opposite spin polarization directions.

\section{Non-resonant ultrafast magnetic x-ray scattering}
In this section, we present the formulae of non-resonant magnetic x-ray scattering (MXS) cross section and the details of ab initio calculation.
We also establish a model for the MXS circular dichroism (CD) signal in the \mol dissociation limit, in order to quantitatively reconstruct the Berry force induced spin separation from MXS CD signal.
Besides, we use the practical parameters of x-ray free electron laser to estimate the detected photon number of MXS CD signal.

\subsection{MXS cross section}
The interaction Hamiltonian between target molecules and incident x-ray is~\cite{BookXrayScatter,Blume57:JAP85}
\begin{eqnarray}
    \hat{H}_{\mathrm{int}}=\sum_j \frac{\alpha^2}{2}\vec{A}(\vec{r}_j)^2 -\alpha\vec{A}(\vec{r}_j)\cdot\hat{\vec{p}}_j-\alpha\hat{\vec{s}}_j\cdot[\nabla\times\vec{A}(\vec{r}_j)]-\frac{\alpha^3}{2}\hat{\vec{s}}_j\cdot[\dot{\vec{A}}(\vec{r}_j)\times\vec{A}(\vec{r}_j)]\,,
\end{eqnarray}
where $\vec{r}_j$, $\hat{\vec{p}}_j$ and $\hat{\vec{s}}_j$ are the position, momentum and spin vectors of the $j$-th electron, $\vec{A}(\vec{r})$ is the vector potential of incident x-ray.
Although the external magnetic field $B=0.1$~T could also contribute to the vector potential $\vec{A}(\vec{r})$, it corresponds to the intensity (in SI units)
\begin{eqnarray}
    I=\frac{1}{2}c\epsilon_0(cB)^2\approx 1.2\times10^{8}\, \mathrm{W}/\mathrm{cm}^2\,,
\end{eqnarray}
which is much weaker compared to incident x-ray and can be ignored.
The differential cross section $\frac{d\sigma}{d\Omega}(\vec{q})$ from initial state $\ket{\psi_i}$ to final state $\ket{\psi_f}$ for x-ray scattering including magnetic scattering term can be derived using Kramers-Heisenberg formula~\cite{BookXrayScatter,Ohsumi1:APX16} as
\begin{eqnarray}\label{Eq:orgdiff}
\frac{d\sigma}{d\Omega}(\vec{q})&=&\alpha^4\left| \BK{\psi_f}{\sum_{j}e^{i\vec{q}\cdot\vec{r}_j}}{\psi_i} (\vec{e}_1\cdot\vec{e}_2^{\,*}) -i\alpha^2 \BK{\psi_f}{\sum_{j}e^{i\vec{q}\cdot\vec{r}_j}(\vec{p}_{j'}\cdot\vec{C}+\vec{\sigma}_{j'}\cdot\vec{D})}{\psi_i} \right|^2\,,
\end{eqnarray}
where $\vec{e}_1$ and $\vec{e}_2$ are the polarization vectors of incident and outgoing x-ray, $\vec{q}$ is momentum transfer of scattered photon, $\alpha^4$ corresponds to the Thomson scattering cross section factor 0.08~barn~\cite{Santra42:PRB09}.
The polarization factors $\vec{C}$ and $\vec{D}$ are
\begin{eqnarray}    \vec{C}&=&\frac{i}{\alpha}(\hat{K}_2-\hat{K}_1)\times(\vec{e}_2^{\,*}\times\vec{e}_1)\,,\\\nonumber
    \vec{D}&=&\frac{\omega}{2}[\vec{e}_2^{\,*}\times\vec{e}_1-(\hat{K}_2\times\vec{e}_2^{\,*})\times(\hat{K}_1\times\vec{e}_1)-(\vec{e}_2^{\,*}\cdot\hat{K}_1)(\hat{K}_1\times\vec{e}_1)+(\vec{e}_1\cdot\hat{K}_2)(\hat{K}_2\times\vec{e}_2^{\,*})]\,,
\end{eqnarray}
where $\hat{K}_1$ and $\hat{K}_2$ are the unit length wavevector of incident and outgoing x-ray, $\omega$ is the photon frequency.
The first term is charge scattering cross section, which reflects charge density distribution of the target molecules.
The second term represents magnetic x-ray scattering signal originated from the interaction between magnetic field of incident x-ray and orbital and spin magnetic dipole moment of the molecule.
In our calculation, the energy of x-ray photon of wavelength $\lambda=0.5$~\AA\xspace is $\sim10^4$~eV, which is much larger compared to the energy level difference of electronic states, which is generally less than 10~eV.

To eliminate the contribution of charge scattering signal and obtain MXS signal, we consider the circular dichroism (CD) signal $\Delta\frac{d\sigma(\vec{q})}{d\Omega}$, which is the difference signal of left- and right-handed circularly polarized x-ray scattering cross section~\cite{BookXrayScatter,Platzman9:PRB70},
\begin{eqnarray}\label{Eq:MXS_formula}
\Delta\frac{d\sigma(\vec{q})}{d\Omega}&=&\frac{d\sigma_\mathrm{L}(\vec{q})}{d\Omega}-\frac{d\sigma_\mathrm{R}(\vec{q})}{d\Omega}=\alpha^6
\operatorname{Re}\left[\BK{\psi}{\hat{\vec{s}}(\vec{q})}{\psi}\cdot\vec{D}(\vec{q})\right]\,,
\end{eqnarray}
where the subscript $\mathrm{L}$ and $\mathrm{R}$ represents left- and right-handed circularly polarized incident x-ray, respectively, $\hat{\vec{s}}(\vec{q})=\sum_{jj'}e^{-i\vec{q}\cdot{(\vec{r}_j-\vec{r}_{j'})}}\hat{\vec{\sigma}}_{j'}$ is the Fourier transformed spin density operator.
If the incident x-ray is along $Z$ axis in lab frame, the polarization vectors for incident x-ray are $\vec{e}_{1\mathrm{L}}=\frac{\sqrt{2}}{2}(1,i,0)^T$ and $\vec{e}_{1\mathrm{R}}=\frac{\sqrt{2}}{2}(1,-i,0)^T$, and then polarization factor $\vec{D}(\vec{q})$ is
\begin{eqnarray}\label{Eq:magvec}
\vec{D}(\vec{q})=-2i(\vec{e}_{1\mathrm{L}}^{\,*}\cdot\vec{e}_{2\mathrm{L}})\vec{D}_\mathrm{L}+2i(\vec{e}_{1\mathrm{R}}^{\,*}\cdot\vec{e}_{2\mathrm{R}})\vec{D}_R=-\omega\begin{pmatrix}
\cos\phi\sin^3\theta\\\sin\phi\sin^3\theta\\\sin^2\theta(1+\cos\theta)
\end{pmatrix}\,.
\end{eqnarray}
%
%
We rewrite the notation of the initial state $\ket{\psi_i}$ as $\ket{\psi}$ for the sake of generality, and summed over all final states $\ket{\psi_f}$, because the ultrafast x-ray scattering signal could cover the final states within the spectral bandwidth of incident ultrashort x-ray pulse~\cite{Santra113:PRL14,Pop8:AS18}.
Based on the MD trajectories, we calculated the MXS signal for the dissociating \mol by
\begin{eqnarray}\label{Eq:MXS_MD}
    \langle\Delta\frac{d\sigma(\vec{q},t)}{d\Omega}\rangle=\frac{\alpha^6}{N_{\mathrm{tr}}}\sum_i\operatorname{Re}\left[\vec{s}^{\,\uparrow\downarrow}_{i}(\vec{q},t)\cdot\vec{D}(\vec{q})\right]\,,
\end{eqnarray}
where $i$ and $N_{\mathrm{tr}}$ are the index and total number of MD trajectories,  $\vec{s}^{\,\uparrow\downarrow}_{i}(\vec{q},t)$, is the spin density contributed from a pair of MD trajectory with opposite spin polarization directions
\begin{eqnarray}\label{Eq:s_pair}
    \vec{s}^{\,\uparrow\downarrow}_{i}(\vec{q},t)=p^{\uparrow}\BK{\psi^{\uparrow}(\vec{R}^{\uparrow}_i,t)}{\hat{\vec{s}}(\vec{q})}{\psi^{\uparrow}(\vec{R}^{\uparrow}_i,t)}+p^{\downarrow}\BK{\psi^{\downarrow}(\vec{R}^{\downarrow}_i,t)}{\hat{\vec{s}}(\vec{q})}{\psi^{\downarrow}(\vec{R}^{\downarrow}_i,t)}\,.
\end{eqnarray}
As shown in Fig.~3 of the main text, the one-dimensional MXS CD cross section is defined by
\begin{eqnarray}\label{Eq:1DMXS}
    \Delta\frac{d\sigma}{d\theta}(q)=\int_0^{2\pi}d\phi \Delta\frac{d\sigma}{d\Omega}(q_x=|\vec{q}|\cos\phi,q_y=|\vec{q}|\sin\phi,q_z=0)\,,
\end{eqnarray}
and we also define the integrated norm of cross section as
\begin{eqnarray}\label{Eq:norm_MXS}
    \Vert\Delta\sigma\Vert&=&\int d\Omega\left|\Delta\frac{d\sigma}{d\Omega}(\vec{q})\right|\\\nonumber
    &=&\int_0^{\theta_{\mathrm{max}}}\sin\theta d\theta \int_0^{2\pi} d\phi \left|\Delta\frac{d\sigma}{d\Omega}(q_x=\frac{2\pi}{\lambda}\sin\theta\cos\phi,q_y=\frac{2\pi}{\lambda}\sin\theta\sin\phi,q_z=0)\right|\,,
\end{eqnarray}
where $\lambda$ is the wavelength of incident x-ray.

\subsection{Ab initio calculation of magnetic x-ray scattering of molecules}

In this section we present the details of ab initio MXS simulation.
The spin density of state $\ket{\psi}$ in the Fourier space is
\begin{eqnarray}\label{Eq:abIni_MXS}
    \BK{\psi}{\hat{\vec{s}}(\vec{q})}{\psi}&=&\BK{\psi}{\sum_{j}\hat{\vec{\sigma}}_{j}}{\psi} +\frac{1}{2} \BK{\psi}{\sum_{j\neq j'}e^{-i\vec{q}\cdot{(\vec{r}_j-\vec{r}_{j'})}}\hat{\vec{\sigma}}_{j'} + e^{-i\vec{q}\cdot{(\vec{r}_{j'}-\vec{r}_{j})}}\hat{\vec{\sigma}}_{j} }{\psi}\\\nonumber
    &=&\int d\vec{x}_1 \hat{\vec{\sigma}}_{1} \gamma(\vec{x}_1^{\,'},\vec{x}_1)+\int d\vec{x}_1 d\vec{x}_2 \left(e^{-i\vec{q}\cdot\vec{r}_{12}}\hat{\vec{\sigma}}_{2} + e^{-i\vec{q}\cdot\vec{r}_{21}}\hat{\vec{\sigma}}_{1}\right)\Gamma(\vec{x}_1^{\,'},\vec{x}_2^{\,'},\vec{x}_1,\vec{x}_2) \,,
\end{eqnarray}
where $\vec{x}=(\vec{r},s)$ includes spatial and spin coordinates, $\gamma(\vec{x}_1^{\,'},\vec{x}_1)$ and $\Gamma(\vec{x}_1^{\,'},\vec{x}_2^{\,'},\vec{x}_1,\vec{x}_2)$ are the one- and two-particle reduced density matrices, respectively.
In Eq.~\ref{Eq:abIni_MXS}, the operators with subscripts 1 and 2 firstly act on the unprimed coordinates $\vec{x}_1$ and $\vec{x}_2$, then the spin density is obtained by taking $\vec{x}_1^{\,'}=\vec{x}_1$ and $\vec{x}_2^{\,'}=\vec{x}_2$.
In molecular orbital (MO) basis, the reduced density matrices can be expanded as
\begin{eqnarray}
\gamma(\vec{x}_1^{\,'},\vec{x}_1)&=&\sum_{kl}\gamma_{lk}\varphi_k^*(\vec{x}_1^{\,'})\varphi_l(\vec{x}_1)\,,\\
\Gamma(\vec{x}_1^{\,'},\vec{x}_2^{\,'},\vec{x}_1,\vec{x}_2)&=&\sum_{klmn}\Gamma_{mnkl}\varphi_k^*(\vec{x}_1^{\,'})\varphi_{l}^*(\vec{x}_2^{\,'})\varphi_m(\vec{x}_1)\varphi_{n}(\vec{x}_2)\,,
\end{eqnarray}
where $\varphi_k(\vec{x})=\varphi_{k_r}(\vec{r})\chi_{k_s}(s)$ is the $k$-th MO with spatial component $\varphi_{k_r}(\vec{r})$ and spin component $\chi_{k_s}(s)$.
The integral of MOs are calculated by
\begin{eqnarray}
    \int d\vec{x}_1  \varphi_k^*(\vec{x}_1)\hat{\vec{\sigma}}_{1}\varphi_l(\vec{x}_1)=\delta_{k_rl_r}\int d\vec{s}_1 \chi^*_{k_s}(s_1)\hat{\vec{\sigma}}_{1}\chi_{l_s}(s_1)\,,
\end{eqnarray}
and
\begin{eqnarray}
    &&\int d\vec{x}_1 d\vec{x}_2 \varphi_k^*(\vec{x}_1)\varphi_{l}^*(\vec{x}_2)\left(e^{-i\vec{q}\cdot\vec{r}_{12}}\hat{\vec{\sigma}}_{2} + e^{-i\vec{q}\cdot\vec{r}_{21}}\hat{\vec{\sigma}}_{1}\right)\varphi_m(\vec{x}_1)\varphi_{n}(\vec{x}_2)\\\nonumber
    &=&\int d\vec{r}_1 d\vec{r}_2 \varphi_{k_r}^*(\vec{r}_1)\varphi_{l_r}^*(\vec{r}_2)   e^{-i\vec{q}\cdot\vec{r}_{12}}\varphi_{m_r}(\vec{r}_1)\varphi_{n_r}(\vec{r}_2)\times\delta_{k_sm_s}\int d\vec{s}_2 \chi^*_{l_s}(s_2)\hat{\vec{\sigma}}_{2}\chi_{n_s}(s_2)\\\nonumber
    &&+\int d\vec{r}_1 d\vec{r}_2 \varphi_{k_r}^*(\vec{r}_1)\varphi_{l_r}^*(\vec{r}_2)   e^{-i\vec{q}\cdot\vec{r}_{21}}\varphi_{m_r}(\vec{r}_1)\varphi_{n_r}(\vec{r}_2)\times\int d\vec{s}_1 \chi^*_{k_s}(s_1)\hat{\vec{\sigma}}_{1}\chi_{m_s}(s_1)\delta_{l_sn_s} \,.
\end{eqnarray}
In the simulation of MXS CD signal, the one- and two-particle reduced density matrices $\gamma_{lk}$ and $\Gamma_{ll'kk'}$ as well as the MO integrals are calculated ab initio.

\subsection{Effect of spin polarized Boltzmann distribution to MXS signal}\label{sec:MXS_noise}

\begin{figure}
    \centering
    \includegraphics[width=10cm]{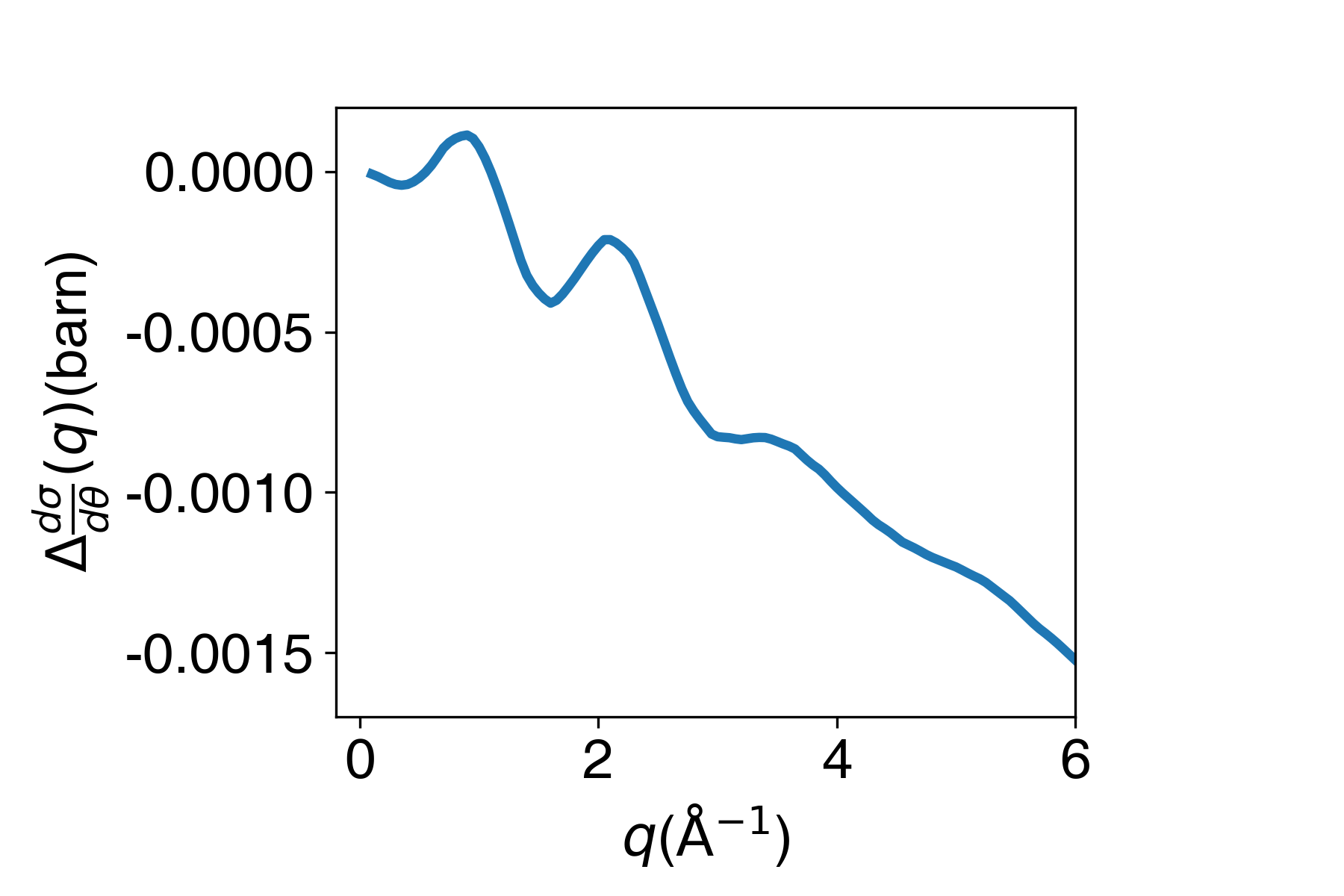}
    \caption{Magnetic x-ray scattering (MXS) signal for a representative \mol geometry shown in Fig.~\ref{fig:MXS_Pr}(a).}
    \label{fig:ab_XMS}
\end{figure}

Based on the MD trajectories, the MXS CD signal of MD trajectories is defined by Eq.~\ref{Eq:MXS_MD} and Eq.~\ref{Eq:s_pair}.
The MXS signal reflecting the effect of Berry force corresponds to $p^{\uparrow}=p^{\downarrow}=\frac{1}{2}$ in Eq.~\ref{Eq:s_pair}.
The spin density signal originated from Boltzmann distribution also lead to the dominant noise, whose magnitude can be estimated by
\begin{eqnarray}
    \Delta\frac{d\sigma^{\mathrm{noise}}(\vec{q},t)}{d\Omega}&\approx& |p^{\uparrow}-p^{\downarrow}| \alpha^6\left\langle\operatorname{Re}[\BK{\psi^{\uparrow}(\vec{R}^{\uparrow}(t))}{\hat{\vec{s}}}{\psi^{\uparrow}(\vec{R}^{\uparrow}(t))}\cdot\vec{D}(\vec{q})]\right\rangle_{\downarrow}\,.
\end{eqnarray}
From Eq.~\ref{Eq:Boltz}, we have $|p^{\uparrow}-p^{\downarrow}|\approx 2\times 10^{-4}$, and $\alpha^6\left\langle\operatorname{Re}[\BK{\psi^{\uparrow}(\vec{R}^{\uparrow}(t))}{\hat{\vec{s}}}{\psi^{\uparrow}(\vec{R}^{\uparrow}(t))}\cdot\vec{D}(\vec{q})]\right\rangle_{\downarrow}$ is the averaged spin density for spin down MD trajectories, whose magnitude can be estimated by the MXS cross section of a representative geometry to be in the order of $10^{-3}~\mathrm{barn}$, as shown in Fig.~\ref{fig:ab_XMS}.
Thus, the noise level $\Delta\frac{d\sigma^{\mathrm{noise}}}{d\Omega}\sim 10^{-7}~\mathrm{barn}$ is by $10^{-2}$ order of magnitude smaller than the MXS CD signal caused by Berry force, which is in the $10^{-5}~\mathrm{barn}$ order of magnitude as shown in Fig.~3(a) in the main text.

\subsection{MXS signal in the dissociation limit}\label{Sec:mxs_diss}
In this section, we elaborate the analytical model for the MXS CD signal, which we established for the purpose to retrieve the spatial spin density evolution in the ultrafast photodissociation dynamics of \mol.
The model is analogous to the independent atom model (IAM) in the theory of molecular diffraction~\cite{Yangjie18:Sci64} and assumes that the spin density is localized.
This assumption becomes rigorous in the dissociation limit of our case.
Because the magnetic field is along $Z$ axis of lab frame, we only consider the $Z$ component of spin density, and the first term of Eq.~\ref{Eq:abIni_MXS} is
\begin{eqnarray}
    \int d\vec{x}_1 \hat{{\sigma}}_{1,Z} \gamma(\vec{x}_1^{\,'},\vec{x}_1)\bigg|_{\vec{x}_1^{\,'}=\vec{x}_1}=\int d\vec{r}_1[\rho(\vec{r}_1\uparrow)-\rho(\vec{r}_1\downarrow)]=\int d\vec{r} s_Z(\vec{r})\,,
\end{eqnarray}
where $\rho(\vec{r}\uparrow)$ and $\rho(\vec{r}\downarrow)$ are spin up and spin down electron densities, respectively,
$s_Z(\vec{r})=\rho(\vec{r}\uparrow)-\rho(\vec{r}\downarrow)$ is the spin density and $\rho(\vec{r})=\rho(\vec{r}\uparrow)+\rho(\vec{r}\downarrow)$ is the spatial density. The term of two-particle reduced density matrix is
\begin{eqnarray}
    &&\int d\vec{x}_1 d\vec{x}_2 e^{-i\vec{q}\cdot\vec{r}_{12}}\hat{{\sigma}}_{2,Z}\Gamma(\vec{x}_1^{\,'},\vec{x}_2^{\,'},\vec{x}_1,\vec{x}_2)\bigg|_{\vec{x}_1^{\,'}=\vec{x}_1,\vec{x}_2^{\,'}=\vec{x}_2}\\\nonumber
    &=&\int d\vec{r}_1 d\vec{r}_2 e^{-i\vec{q}\cdot\vec{r}_{12}} [\rho^{(2)}(\vec{r}_1\uparrow,\vec{r}_2\uparrow)+\rho^{(2)}(\vec{r}_1\downarrow,\vec{r}_2\uparrow)-\rho^{(2)}(\vec{r}_1\uparrow,\vec{r}_2\downarrow)-\rho^{(2)}(\vec{r}_1\downarrow,\vec{r}_2\downarrow)]\\\nonumber
    &=&\frac{N-1}{2N}\int d\vec{r}_1 d\vec{r}_2  e^{-i\vec{q}\cdot\vec{r}_{12}}[\rho(\vec{r}_1\uparrow)+\rho(\vec{r}_1\downarrow)]
    [\rho(\vec{r}_2\uparrow)-\rho(\vec{r}_2\downarrow)]\\\nonumber
    &=&\frac{N-1}{2N}\int d\vec{r}_1 d\vec{r}_2
    e^{-i\vec{q}\cdot\vec{r}_{12}} \rho(\vec{r}_1) s_Z(\vec{r}_2)\,,
\end{eqnarray}
where we neglected the electronic correlation effects and suppose that $\rho^{(2)}(\vec{x}_1,\vec{x}_2)=\frac{N-1}{2N}\rho(\vec{x}_1)\rho(\vec{x}_2)$, and $N$ is the number of electrons.
Thus, the spin density of the state $\ket{\psi}$ is
\begin{eqnarray}\label{Eq:psi_sz}
    s_Z(\vec{q})&=&\BK{\psi}{\hat{s}_Z(\vec{q})}{\psi}\\\nonumber
    &=&\int d\vec{r} s_Z(\vec{r})+\frac{N-1}{2N}\int d\vec{r}_1 d\vec{r}_2
    [e^{-i\vec{q}\cdot\vec{r}_{12}} \rho(\vec{r}_1) s_Z(\vec{r}_2)+e^{-i\vec{q}\cdot\vec{r}_{21}} s_Z(\vec{r}_1) \rho(\vec{r}_2)]\\\nonumber
    &=&\int d\vec{r} s_Z(\vec{r})+\frac{N-1}{N}\int d\vec{r}_1 d\vec{r}_2
    e^{-i\vec{q}\cdot\vec{r}_{21}} s_Z(\vec{r}_1) \rho(\vec{r}_2)\,.
\end{eqnarray}

\begin{figure}
    \centering
    \includegraphics[width=8cm]{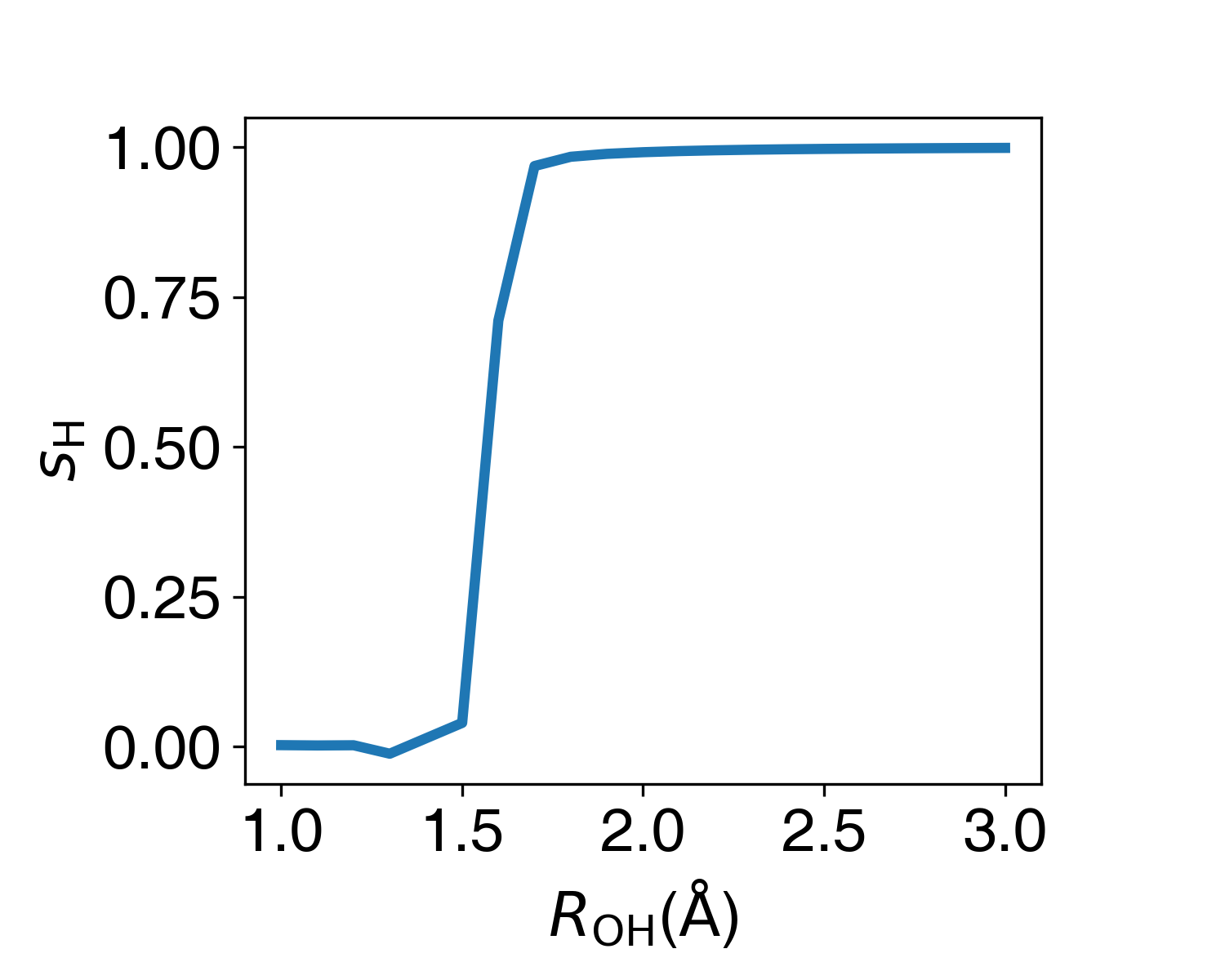}
    \caption{The change of spin density of dissociated H atom $s_\mathrm{H}$ with O--H distance $R_{\mathrm{OH}}$ obtained by Mulliken population analysis.
    Around equilibrium geometry in the ground state, the spin density is mainly distributed on C atom.
    When $R_{\mathrm{OH}}>2$~\AA, spin density is almost completely localized on the dissociated H atom with $s_{\mathrm{H}}$ approaching 1.}
    \label{fig:diss_sH}
\end{figure}

Fig.~\ref{fig:diss_sH} shows the change of spin density $s_\mathrm{H}$ of the H atom with O--H distance $R_{\mathrm{OH}}$.
The Mulliken population analysis of spin density is ab initio calculated using Molpro package~\cite{MOLPRO2019} on the RASSCF(5,7)/cc-pvtz level.
In the dissociation of \mol, when $R_{\mathrm{OH}}>2$~\AA, $s_\mathrm{H}\approx 1$ and the spin density is completely localized on the dissociated H atom.
We assume the spin density to be spherical symmetric
\begin{eqnarray}
    s_Z(\vec{r})=\pm\rho_{\mathrm{H}}(|\vec{r}-\vec{r}_{\mathrm{H}}|)\,,
\end{eqnarray}
where positive and negative sign corresponds to spin up and spin down molecules, respectively.
The total electron density is contributed from both CH$_2$O and the dissociated H atom
\begin{eqnarray}
    \rho(\vec{r})=\rho_{X}(|\vec{r}-\vec{r}_{X}|)+\rho_{\mathrm{H}}(|\vec{r}-\vec{r}_{\mathrm{H}}|)\,,
\end{eqnarray}
where $X$ represents the $\mathrm{CH}_2\mathrm{O}$ moiety, it is regarded as a pseudoatom in the model, so Eq.~\ref{Eq:psi_sz} can be expressed as
\begin{eqnarray}
    s_Z(\vec{q})
    &=&\pm N_{\mathrm{un}}\pm\frac{N-1}{N}e^{i\vec{q}\cdot{\vec{r}_{\mathrm{H}}}}\int d\vec{r}_1 e^{i\vec{q}\cdot{\vec{r}_{1}}} \rho_{\mathrm{H}}(r_1)\\\nonumber
    &&\times\left[e^{-i\vec{q}\cdot{\vec{r}_{X}}} \int d\vec{r}_2 e^{-i\vec{q}\cdot{\vec{r}_{2}}}\rho_{X}(r_2)+e^{-i\vec{q}\cdot{\vec{r}_{\mathrm{H}}}} \int d\vec{r}_2 e^{-i\vec{q}\cdot{\vec{r}_{2}}}\rho_{\mathrm{H}}({r}_2)\right]\\\nonumber
    &=&\pm N_{\mathrm{un}}\pm\frac{N-1}{N}\left[f^2_{\mathrm{H}}(q)+e^{i\vec{q}\cdot({\vec{r}_{\mathrm{H}}}-\vec{r}_{X})}f_{\mathrm{H}}(q)f_{X}(q)\right]\,,
\end{eqnarray}
where $N_{\mathrm{un}}$ is the number of unpaired electrons in \mol, and
\begin{eqnarray}
f_{i}(q)=\int d\vec{r}e^{i\vec{q}\cdot{\vec{r}}} \rho_{i}(r)=\int_0^\infty dr 4\pi r^2\rho_{i}(r)\frac{\sin (qr)}{qr},\quad (i=\mathrm{H},X)
\end{eqnarray}
is the form factor for H atom or CH$_2$O moiety.
Assume the isotropic rotational wavepacket of gas phase \mol, the one-dimensional MXS CD cross section is
\begin{eqnarray}\label{Eq:diss_MXS}
    \Delta \frac{d\sigma}{d\theta}(q)&=&2\pi\alpha^6 D_Z(q) \langle s_Z(q)\rangle_{\mathrm{rot}}\,,\\\label{Eq:diss_MXS_2}
    \langle s_Z(q)\rangle_{\mathrm{rot}}&=& \pm N_{\mathrm{un}}\pm\frac{N-1}{N}[f^2_{\mathrm{H}}(q)+\frac{\sin(qR_{\mathrm{OH}})}{qR_{\mathrm{OH}}}f_{\mathrm{H}}(q)f_{X}(q)]\,.
\end{eqnarray}
Note that Eq.~\ref{Eq:diss_MXS} and Eq.~\ref{Eq:diss_MXS_2} are similar to the rotational averaged charge scattering x-ray diffraction signal in the independent atom model~\cite{Budarz49:JPB16}
\begin{eqnarray}\label{Eq:XD_IAM}
    I_{\mathrm{IAM}}(q)\propto\sum_I |f_I(q)|^2+\sum_{I\neq J} f_I(q)f_J(q)\frac{\sin(q r_{IJ})}{q r_{IJ}}\,,
\end{eqnarray}
where $I,J$ are atomic indexes, $f_I(q)$ is the form factor of the $I$-th atom, and $r_{IJ}$ is the distance between the $I$-th atom and $J$-th atom.
The second term in Eq.~\ref{Eq:XD_IAM} reflects molecular structural information, which can be interpreted as the superposition of interference pattern contributed by each pair of atoms.
Likewise, the MXS CD signal can be regarded as the interference pattern by the CH$_2$O moiety and dissociated H atom, and the spacing of fringes is $\Delta q\sim\frac{1}{R_{\mathrm{OH}}}$, due to the factor $\frac{\sin(qR_{\mathrm{OH}})}{qR_{\mathrm{OH}}}$.
$\Delta q$ decreases with the dissociation of \mol molecule, and goes beyond resolution with \mol dissociation, which explains the descent of MXS CD signal intensity as shown in Fig.~3(a) in the main text.
To quantitatively interpret the change of signal intensity with $R_{\mathrm{OH}}$, we calculated the MXS CD signal contributed by the average O--H distance for spin up and spin down MD trajectories $\langle R^\uparrow_{\mathrm{OH}} \rangle$ and $\langle R^\downarrow_{\mathrm{OH}} \rangle$.
The integrated norm $\Vert\Delta\sigma^{\mathrm{ave}}\Vert$ of the MXS CD signal, as defined in Eq.~\ref{Eq:norm_MXS}, decreases after 200~fs with the \mol dissociation.

\begin{figure}
    \centering
    \includegraphics[width=6cm]{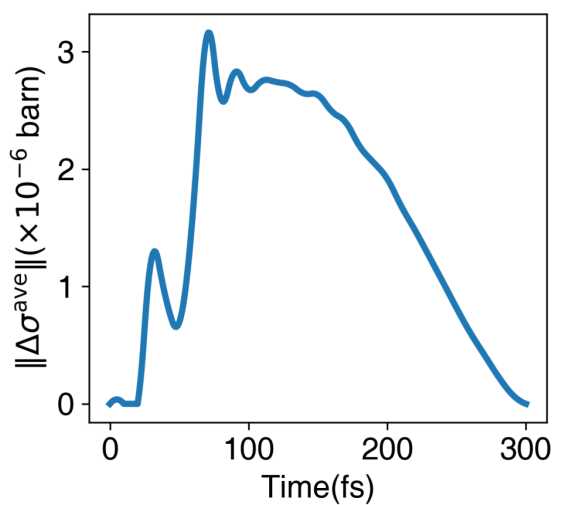}
    \caption{Temporal evolution for integrated norm of cross section $\Vert\Delta\sigma^{\mathrm{ave}}\Vert$ of MXS CD signal calculated using average O--H distance for spin up and spin down MD trajectories $\langle R^\uparrow_{\mathrm{OH}} \rangle$ and $\langle R^\downarrow_{\mathrm{OH}} \rangle$.
    The spacing of diffraction streaks $\Delta q\sim\frac{1}{R_{\mathrm{OH}}}$ decrease with the dissociation of \mol, which goes beyond resolution with \mol dissociation and leads to the descent of MXS CD signal intensity after 200~fs.
    }
    \label{fig:MXS_ave_norm}
\end{figure}

Eq.~\ref{Eq:diss_MXS} can be analytically reversed to solve $R_{\mathrm{OH}}$ from MXS CD signal, using the following orthogonal property
\begin{eqnarray}
    &&\int_{0}^{+\infty} dq q^2\frac{\sin(qR_{\mathrm{OH}})}{qR_{\mathrm{OH}}} \frac{\sin(qR'_{\mathrm{OH}})}{qR'_{\mathrm{OH}}}\\\nonumber
    &=&\frac{1}{2R_{\mathrm{OH}}R'_{\mathrm{OH}}} \int_{0}^{+\infty} dq[\cos(qR_{\mathrm{OH}}-qR'_{\mathrm{OH}})-\cos(qR_{\mathrm{OH}}+qR'_{\mathrm{OH}})]\\\nonumber
    &=&\frac{\pi}{2R^2_{\mathrm{OH}}}\delta(R_{\mathrm{OH}}-R'_{\mathrm{OH}})\,,\quad (R_{\mathrm{OH}}>0,R'_{\mathrm{OH}}>0)\,,
\end{eqnarray}
where we have used the formula
\begin{eqnarray}
    \int_0^{+\infty} dq \cos(qx)=\frac{1}{2} \int_{-\infty}^{+\infty} dq \cos(qx)=\frac{1}{4} \int_{-\infty}^{+\infty} dq (e^{iqx}+e^{-iqx})=\pi \delta(x)\,.
\end{eqnarray}
So the O--H distance can be retrieved from Eq.~\ref{Eq:diss_MXS} by
\begin{eqnarray}
    \frac{2R^2}{\pi}\int_{0}^{+\infty} dq q^2\frac{\sin(qR)}{qR} \left[\frac{N}{N-1}\frac{\langle s_Z(q)\rangle_{\mathrm{rot}}-N_{\mathrm{un}}}{f_{\mathrm{H}}(q)f_{X}(q)}-\frac{f_{\mathrm{H}}(q)}{f_X(q)}\right]=\delta(R-R_{\mathrm{OH}})\,.
\end{eqnarray}

\begin{figure}
    \centering
    \includegraphics[width=15cm]{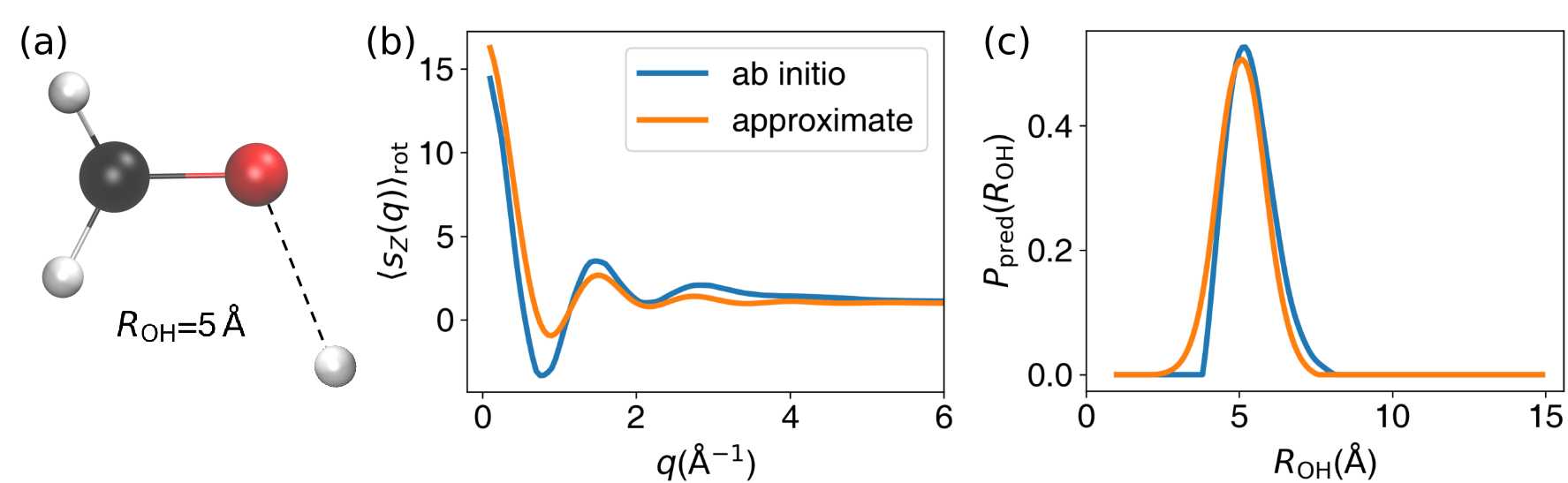}
    \caption{Rotational averaged spin density for a representative dissociated geometry and retrieved probability density of O--H distance.
    (a) The representative dissociated geometry with O--H distance $R_{\mathrm{OH}}=5$~\AA.
    Black, red and white spheres represent C, O and H atoms, respectively.
    (b) The blue line is the rotational averaged spin density obtained from ab initio calculated MXS CD cross section $\langle s_Z(q)\rangle_{\mathrm{rot}}=\frac{1}{2\pi\alpha^6 D_Z(q)}\frac{d\sigma}{d\theta}$ from Eq.~\ref{Eq:diss_MXS}, and the orange line is calculated by the analytical model derived in Eq.~\ref{Eq:diss_MXS_2}.
    (c) Retrieved probability density of $R_{\mathrm{OH}}$ centered at    $R_{\mathrm{OH}}=5$~\AA.
    }
    \label{fig:MXS_Pr}
\end{figure}

We choose a representative dissociated geometry in Fig.~\ref{fig:MXS_Pr}(a) with $R_{\mathrm{OH}}=5$~\AA\xspace to validate the analytical model we made in this section to calculate the MXS CD signal and spin density in the dissociation limit.
As shown in Fig.~\ref{fig:MXS_Pr}(b), the rotational averaged spin density calculated from Eq.~\ref{Eq:diss_MXS_2} is consistent with the ab initio calculated result, and the retrieved probability density of O--H distance is centered at $R_{\mathrm{OH}}=5$~\AA\xspace as shown in Fig.~\ref{fig:MXS_Pr}(c).

The retrieval method can be generalized to obtain the distribution of $R_{\mathrm{OH}}$ in the swarm of MD trajectories.
The distribution of O--H distance for all MD trajectories are characterized by the spin-dependent probability density function $P^\uparrow(R_{\mathrm{OH}},t)$ and $P^\downarrow(R_{\mathrm{OH}},t)$.
After averaging over all MD trajectories, the MXS CD signal is
\begin{eqnarray}
    \langle\Delta \frac{d\sigma}{d\theta}(q,t)\rangle =2\pi\alpha^6 D_Z \frac{N-1}{N} f_{\mathrm{H}}f_{X}\int_0^{+\infty} dR_{\mathrm{OH}}[P^\uparrow(R_{\mathrm{OH}},t)-P^\downarrow(R_{\mathrm{OH}},t)] \frac{\sin(qR_{\mathrm{OH}})}{qR_{\mathrm{OH}}}\,.
\end{eqnarray}
Define $A(q)=\frac{N-1}{Nq}\pi^2\alpha^6 D_Z(q)  f_{\mathrm{H}}(q)f_{X}(q)$ and $\Delta P(R_{\mathrm{OH}},t)=P^\uparrow(R_{\mathrm{OH}},t)-P^\downarrow(R_{\mathrm{OH}},t)$, the equation can be expressed as
\begin{eqnarray}
    \langle\Delta \frac{d\sigma}{d\theta}(q,t)\rangle=\frac{2q}{\pi}A(q) \int_0^{+\infty} dR_{\mathrm{OH}} \Delta P(R_{\mathrm{OH}},t) \frac{\sin(qR_{\mathrm{OH}})}{qR_{\mathrm{OH}}}\,.
\end{eqnarray}
The spatially separated spin distribution, which is given as a function of O--H distance, can be retrieved from the MXS CD intensity by
\begin{eqnarray}
    &&\int_{0}^{+\infty} dq \frac{q^2}{\frac{2q}{\pi}A(q)} \frac{\sin(qR'_{\mathrm{OH}})}{qR'_{\mathrm{OH}}} \langle\Delta \frac{d\sigma}{d\theta}(q,t)\rangle \\\nonumber
    &=& \int_0^{+\infty} dR_{\mathrm{OH}} \Delta P(R_{\mathrm{OH}},t) \int_{0}^{+\infty} dq q^2\frac{\sin(qR_{\mathrm{OH}})}{qR_{\mathrm{OH}}} \frac{\sin(qR'_{\mathrm{OH}})}{qR'_{\mathrm{OH}}}\\\nonumber
    &=&\int_0^{+\infty} dR_{\mathrm{OH}} \Delta P(R_{\mathrm{OH}},t) \delta(R_{\mathrm{OH}}-R'_{\mathrm{OH}})\frac{\pi}{2R'^2_{\mathrm{OH}}}\\\nonumber
    &=& \Delta P(R'_{\mathrm{OH}},t)\frac{\pi}{2R'^2_{\mathrm{OH}}}\,,
\end{eqnarray}
so
\begin{eqnarray}\label{Eq:PrRt}
    \Delta P(R_{\mathrm{OH}},t)=R_{\mathrm{OH}} \int_{0}^{+\infty} dq \frac{\sin(qR_{\mathrm{OH}})}{A(q)} \langle\Delta \frac{d\sigma}{d\theta}(q,t)\rangle\,.
\end{eqnarray}

\subsection{Estimate of the scattered x-ray photon number}
In this section, we use the practical parameters of x-ray free electron laser (XFEL)~\cite{Emma4:NP10} to estimate the detected photon number of magnetic x-ray scattering (MXS). The total number of scattered photons $N_{\mathrm{total}}$ is
\begin{eqnarray}
N_{\mathrm{total}}=N_{\mathrm{MXS}}N_{\mathrm{pulse}}N_{\mathrm{mol}}\,,
\end{eqnarray}
where $N_{\mathrm{MXS}}$ is the number of scattered photons for one incident XFEL pulse and one target molecule, $N_{\mathrm{pulse}}$ is the number of incident XFEL pulses and $N_{\mathrm{mol}}$ is the number of target molecules. First, we have
\begin{eqnarray}
N_{\mathrm{MXS}}&=&\sigma\frac{N_{\gamma}}{d^2}\,,
\end{eqnarray}
where $\sigma$ is the MXS cross section, $N_{\gamma}$ is the number of photons per pulse, $d$ is the photon source size, and
\begin{eqnarray}
N_{\mathrm{mol}}=\rho d^2L\,,
\end{eqnarray}
where $\rho$ is the molecular gas density, $d^2L$ is the interaction volume and $L$ is the diameter of molecular gas jet, and
\begin{eqnarray}
N_{\mathrm{pulse}}=\nu T\,,
\end{eqnarray}
where $\nu$ is the repetition rate and $T$ is the time of measurement. Using the parameters of XFEL, the scattered photon number per second for $\sigma=10^{-5}$~barn (see Fig.~3(a) in the main text) is
\begin{eqnarray}
N_{\mathrm{MXS}}&=&\sigma N_{\gamma}\rho L \nu T \\\nonumber
&\approx&10^{-5}~\mathrm{barn}\times10^{12}\times10^{16}~\mathrm{cm}^{-3}\times 1~\mathrm{cm}\times 1~\mathrm{MHz} \times 1~\mathrm{s}=10^{5}\,.
\end{eqnarray}


%